\documentclass[submission,copyright,creativecommons]{eptcs}
\providecommand{\event}{GCM 2019} % Name of the event you are submitting to
\usepackage{bm}
\usepackage[utf8]{inputenc}
\usepackage{fp}
\usepackage{pifont}
\usepackage{tikz}                     %% Portable Bilder 
 \usetikzlibrary{matrix,chains,graphs,backgrounds,positioning,fit,decorations.pathmorphing,shapes,arrows,calc,shadows}
\usepackage{graphicx}
\usepackage{amstext}
\usepackage{amsmath}
\usepackage{amssymb}
\usepackage{bm}
\usepackage{pgfplots}

\newtheorem{defn}{Definition}
\newtheorem{example}{Example}

\newcommand{\figref}[1]{\hyperref[#1]{Fig.~\ref*{#1}}}
\newcommand{\tabref}[1]{\hyperref[#1]{Table~\ref*{#1}}}
\newcommand{\sectref}[1]{\hyperref[#1]{Sect.~\ref*{#1}}}
\newcommand{\sectionref}[1]{\hyperref[#1]{Section~\ref*{#1}}}
\newcommand{\defref}[1]{\hyperref[#1]{Def.~\ref*{#1}}}
\newcommand{\thmref}[1]{\hyperref[#1]{Thm.~\ref*{#1}}}
\newcommand{\propref}[1]{\hyperref[#1]{Prop.~\ref*{#1}}}
\newcommand{\lemmaref}[1]{\hyperref[#1]{Lemma~\ref*{#1}}}
\newcommand{\exref}[1]{\hyperref[#1]{Example~\ref*{#1}}}
\newcommand{\tup}[1]{\langle #1 \rangle}   % tuple <x,y>
\newcommand{\Lab}{\Sigma}  % (constant) labels
\newcommand{\TLab}{\mathcal{T}}  % terminals
\newcommand{\NLab}{\mathcal{N}}  % nonterminals
\newcommand{\arity}{\mathit{arity}}   % attachments of variables
\newcommand{\Nat}{\mathbb{N}}

\renewcommand{\L}{{\mathcal{L}}}

\newcommand{\R}{\mathcal{R}}

\newcommand{\To}{\Rightarrow}
\newcommand{\too}{\displaystyle\operatornamewithlimits{\to}}
\newcommand{\Too}{\mathop{\displaystyle\operatornamewithlimits{\Rightarrow}}}
\newcommand{\circled}[1]{{\FPeval{\result}{clip(#1+171)}\ding{\result}}}

\def\literal#1{\bm{#1}}
\def\der{\mathop{\Rightarrow}}
\def\epsilon{\varepsilon}
\def\emptyseq{\varepsilon}        % empty sequence
\def\emptyset{\varnothing}
\def\qrule#1{(#1)}
\def\startSym{\ntlabel{Z}}
\def\startLit{\startSym^\emptyseq}
\def\graph(#1,#2){\langle {#1}, {#2}\rangle}
\def\Gr#1{\mathcal{G}_{#1}}

\def\perm{\bowtie}
\def\ntlabel#1{\mathsf{#1}}
\def\node#1{\setcounter{varctr}{#1}\mathfrak{\alph{varctr}}}
\def\tlit#1{\lb t^{#1}}
\def\Dlit#1{\lb D^{#1}}
\def\triangle(#1,#2,#3){\lb t^{\node{#1}\node{#2}\node{#3}}}
\def\nt(#1,#2,#3){\lb D^{\node{#1}\node{#2}\node{#3}}}
\def\state#1#2#3{#1^{\mathit{#2}}_{\mathrm{#3}}}
\pgfplotsset{
        width=7cm,
        scaled ticks = false,
        grid = major,
        xmin=0,
        ymin=0,
        legend style={cells={anchor=west}},
    	tick label style={font=\fontsize{7}{8}\selectfont},
    	title style={font=\fontsize{8}{9}\selectfont},
    	label style={font=\fontsize{7}{8}\selectfont},
    	legend style={font=\fontsize{7}{8}\selectfont},
}

\tikzset{%
  o/.style={circle,draw,fill=white,inner sep=0.0pt,minimum size=2mm},
}

\newcounter{varctr}
\title{Speeding up Generalized PSR Parsers \\by Memoization Techniques}
\author{Mark Minas
  \institute{Universität der Bundeswehr München\\Neubiberg, Germany}
  \email{mark.minas@unibw.de}
}
\def\titlerunning{Speeding up Generalized PSR Parsers by Memoization Techniques}
\def\authorrunning{Mark Minas}
\begin{document}
\maketitle
\begin{abstract}
  Predictive shift-reduce (PSR) parsing for hyperedge replacement (HR)
grammars is very efficient, but restricted to a subclass of unambiguous
HR grammars. To overcome this restriction, we have recently extended
PSR parsing to generalized PSR (GPSR) parsing along the lines of
Tomita-style generalized LR parsing. Unfortunately, GPSR parsers turned out
to be too inefficient without manual tuning. This paper proposes to use
memoization techniques to speed up GPSR parsers without any need of
manual tuning, and which has been realized within the graph parser
distiller \textsc{Grappa}. We present running time measurements for
some example languages; they show a significant speed up by some orders
of magnitude when parsing valid graphs. But memoization techniques do
not help when parsing invalid graphs or if all parses of an ambiguous
input graph shall be determined.

\end{abstract}
% \keywords{hyperedge replacement grammar, graph parsing, parser generator}

% !TEX root = main.tex

\section{Introduction}\label{s:intro}

In earlier work \cite{drewes-hoffmann-minas-18}, we have devised
\emph{predictive shift-reduce} parsing (PSR), which lifts D.E.\ Knuth's
LR string parsing \cite{Knuth:65} to graphs and runs in at most
expected linear time in the size of the input graph. However, parsing
for graph grammars based on hyperedge replacement (HR) is in general
NP-hard, even for a particular grammar~\cite[sect.~7.1]{Drewes-Habel-Kreowski:97}. Therefore, PSR
parsing is restricted to a subclass of HR grammars, which particularly
must be unambiguous. We have recently extended PSR parsing to
\emph{generalized PSR} (GPSR) parsing~\cite{hoffmann-minas:19}, which
can be applied to every HR grammar.

GPSR parsing has been motivated by \emph{generalized LR}
(GLR) parsing for strings, originally devised by
M.~Tomita~\cite{Tomita:85}, and extended and improved by several
authors (for an overview see~\cite{Scott:06}). The original GLR parsing
algorithm by Tomita runs in $O(n^{k+1})$ where $k$ is the length of the longest rule, whereas
improved versions like \emph{Binary Right Nulled GLR} (BRNGLR) parsers run in
worst-case cubic time~\cite{Scott:07}.

GPSR parsing cannot be efficient in general because GPSR parsers can be
applied to every HR grammar. But our
experiments~\cite{hoffmann-minas:19} have shown that GPSR parsers are
even slower than simple graph parsers that extend the \emph{Cocke-Younger-Kasami} (CYK) algorithm to
graphs~\cite{lautemann:90,Minas:01a}. Manual tuning of GPSR parsers by
using language specific strategies (see \sectref{s:gpsr}) helped to
improve their efficiency, but even those tailored parsers have
not always been faster than the corresponding CYK parsers.

GPSR parsers identify parses of an input graph in a search process that
may run into dead ends. They are inefficient because they waste time in
this process and because they discard all information collected in
these dead ends, even if it could be used later. This paper proposes to
use memoization techniques to keep the information and to reuse it later.
Reuse allows to skip long sequences of parsing operations that would
just recreate information that has already been collected earlier. 

GPSR parsing with memoization has been implemented in the graph-parser distiller \textsc{Grappa}%
\footnote{Available under
  \url{www.unibw.de/inf2/grappa}.\label{fn:grappa}}.
Experiments with generated parsers for different example languages demonstrate that memoization substantially improves parsing speed.

Note that defining graph languages by graph grammars and using graph
parsing is not the only way to check the validity of graphs.
A different widespread approach (e.g.,~\cite{Taentzer-Rensink:05}) is
to use meta-models with additional constraints (e.g., OCL constraints).
Checking whether a graph conforms to a given meta-model and the
constraints can be easier than graph parsing. But it is generally
considered more difficult to come up with a complete set of constraints
that are accepted by all valid graphs, but violated by all invalid
graphs. For instance, defining just the set of all series-parallel
graphs or all flowchart graphs by constraints is non-trivial, but
straight-forward with graph grammars (see~\sectref{s:eval}). In those
cases, efficient graph parsing may be favored over meta-model and
constraint checking.

The remainder of this paper is structured as follows. After recalling
HR grammars in \sectref{s:hr}, PSR parsing in \sectref{s:psr}, and GPSR
parsing in \sectref{s:gpsr}, we describe how memoization can speed up
GPSR parsing in \sectref{s:memo}. We compare its performance with plain
GPSR parsing and with CYK parsing using three example graph languages 
in \sectref{s:eval}:
Sierpinski graphs, series-parallel graphs, and structured flowcharts,
where GPSR parsing with memoization substantially improves plain GPSR
parsing. \sectref{s:concl} concludes the paper.

% !TEX root = main.tex

\section{Graph Grammars Based on Hyperedge Replacement}\label{s:hr}

\def\lb#1{\textsf{#1}}
\def\Ed{{\mathcal E}_{\Lab}}
\def\EdN{{\mathcal E}_{\NLab}}
\def\lab(#1){\ell(#1)}
\def\Vx#1{V_{\!#1}}
\def\startGraph{Z}
Throughout the paper, we assume that $X$ is a global, countably
infinite supply of \emph{nodes}, and that $\Lab$ is a finite set of
\emph{symbols} that comes with an \emph{arity function}
$\arity\colon \Lab \to \Nat$, and is partitioned into disjoint subsets
$\NLab$ of \emph{nonterminals} and and $\TLab$ of \emph{terminals}.

We write hyperedges with their attached nodes as literals and
hypergraphs as ordered sequences of literals. This first may seem
unusual, but it will turn out to be beneficial as parsers will read
hyperedges of the input hypergraph in a certain order.
\begin{defn}[Hypergraph]\label{def:graphs}
  For a symbol $\lb a \in \Lab$ and $k=\arity(\lb a)$ pairwise
  distinct nodes $x_1,\dots,x_k\in X$, a \emph{literal}
  $\literal a = \lb a^{x_1 \cdots x_k} $
  is a \emph{hyperedge} that is labeled with $\lb a$ and
  attached to $x_1, \dots, x_k$. $\Ed$ denotes the set of all literals
  (over $\Lab$).

  A \emph{hypergraph $\gamma = \graph(V, \phi)$ over $\Lab$} consists of a finite set 
  $V \subseteq X$ of nodes and a sequence
  $\phi = \literal e_1 \cdots \literal e_n\in\Ed^*$ of literals such that all
  nodes in these literals are in~$V$. $\Gr{\Lab}$ denotes the set of all hypergraphs over ${\Lab}$. 

  We say that two hypergraphs $\gamma = \graph(V, \phi)$ and 
  $\gamma' = \graph(V', \phi')$ are \emph{equivalent}, written
  $\gamma \perm \gamma'$, if $V = V'$ and $\phi$ is a permutation 
  of~$\phi'$.
\end{defn}
In the following, we usually call hypergraphs just \emph{graphs} and 
hyperedges just \emph{edges} or \emph{literals}. For a graph $\gamma = \graph(V, \phi)$, 
we use the notation $V_\gamma = V$.

Note that a graph $\graph(V, \phi)$ may contain the same literal more
than once in $\phi$, representing indistinguishable, i.e., parallel
edges. Note also that graphs are sequences rather than multisets of
literals, i.e., two graphs
$\graph(V, \phi)$ and $\graph(V', \phi')$ with the same set of nodes, but 
with different sequences of literals are considered to differ, even if $V=V'$ and $\phi'$ 
is just a permutation of $\phi$. However, such graphs are equivalent, denoted 
by the equivalence relation~$\perm$. In contrast, ``ordinary'' graphs would rather be 
represented using multisets of literals instead of 
sequences. The equivalence classes of graphs, 
therefore, correspond to conventional graphs. The ordering of literals 
is technically convenient for the constructions in this paper. However, input graphs to be
parsed should of course be considered up to equivalence. Thus, we will make
sure that the developed parsers yield identical results on graphs $g,g'$ with $g\perm g'$. 

An injective function
$\rho\colon X \to X$ is called a \emph{renaming}, and $\gamma^\rho$ denotes 
the graph obtained by replacing all
nodes in $\gamma$ according to $\rho$. Although renamings are, for technical 
simplicity, defined as functions on the whole of $X$, only the finite subset 
$V_\gamma \subseteq X$ will be relevant. 
We define the ``concatenation'' of two graphs 
$\gamma = \graph(V,\phi), \gamma' =\graph(V',\phi') \in \Gr\Lab$ as $\gamma \gamma'=\graph(V \cup V', \phi \phi')$. 
If a graph $\gamma = \graph(V,\phi)$ is completely determined by its sequence 
$\phi$ of literals, i.e., if each node in $V$ also occurs 
in some literal in $\phi$, we simply use $\phi$ as a shorthand 
for~$\gamma$. 
In particular, a literal $\literal a = \lb a^{x_1 \cdots x_k} \in \Ed$
is identified with the graph $\graph(\{x_1, \ldots, x_k\}, \literal
a)$.

A \emph{hyperedge replacement rule} $r=\qrule{\literal A \to \alpha}$
(\emph{rule} for short) has a nonterminal edge $\literal A \in \EdN$ as its
\emph{left-hand side}, and a graph $\alpha \in \Gr{\Lab}$ with
$\Vx{\literal A} \subseteq \Vx\alpha$ as its \emph{right-hand side}.

Consider a graph $\gamma = \beta \bar{\literal A} \bar\beta \in \Gr\Lab$ with a
nonterminal edge $\bar{\literal A}$ and a rule $r=\qrule{\literal A \to \alpha}$.
A renaming $\mu\colon X \to X$ is a \emph{match} (of $r$ in $\gamma$)
if $\literal A^\mu = \bar{\literal A}$ and if
$\Vx\gamma \cap \Vx{\alpha^\mu} \subseteq \Vx{\literal A^\mu}$.%
\footnote{I.e., a match $\mu$ makes sure that the nodes of
  $\alpha^\mu$ that do not occur in $\bar{\literal A} = \literal A^\mu$ do not collide
  with the other nodes in $\gamma$.} %
A match $\mu$ of $r$ \emph{derives} $\gamma$ to the graph
$\gamma' = \beta \alpha^\mu \bar\beta$.  This is denoted as
$\gamma \der_{r,\mu} \gamma'$. If $\R$ is a finite set of rules, we write $\gamma \To_\R \gamma'$ if $\gamma \der_{r,\mu} \gamma'$ for some match $\mu$ of some rule $r \in \R$.

\begin{defn}[HR Grammar]\label{def:hr-grammar}
  A \emph{hyperedge replacement grammar}
  $\Gamma = (\Lab,\TLab,\R,\startGraph)$ (\emph{HR grammar} for short)
  consists of \emph{symbols} $\Lab$ with \emph{terminals}
  $\TLab \subseteq \Lab$ as assumed above, a finite set $\R$ of rules,
  and a \emph{start graph}~$\startGraph = \startLit$ with
  $\startSym \in \NLab$ of arity $0$.
  $\Gamma$ generates the language
  $\L(\Gamma) = \{ g \in \Gr{\TLab} \mid
                 \startGraph \der\nolimits_\R^* g \}.
  $
\end{defn}

In the following, 
we simply write $\der$ and $\der^*$  because the rule set $\R$ 
in question will always be clear from the context.

\begin{figure}[bt]
	\centering
	\includegraphics[width=0.7\textwidth]{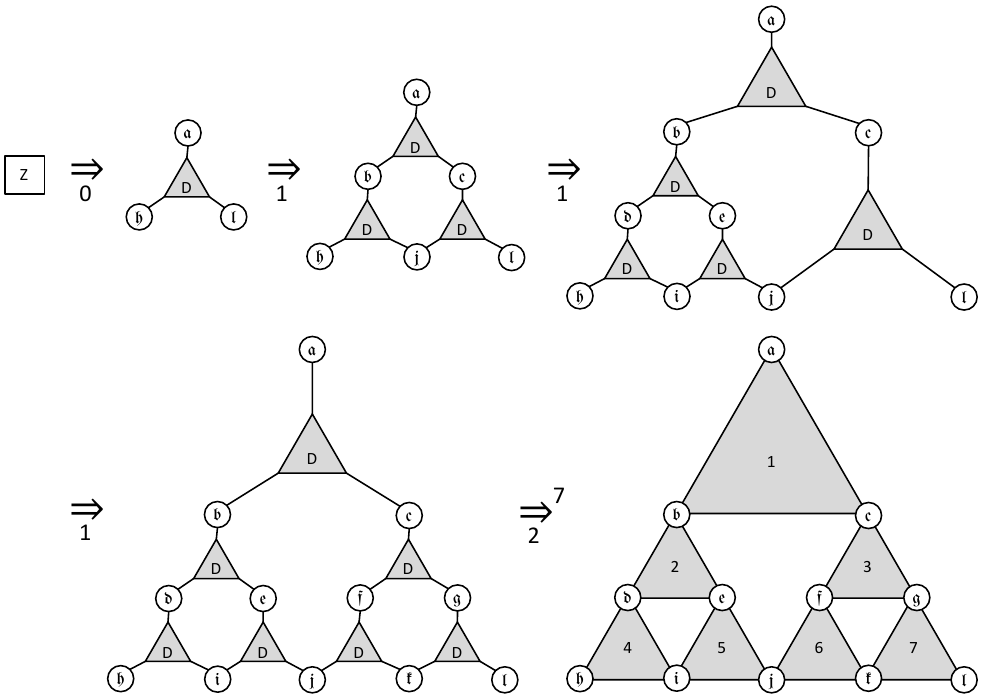}
	\caption{A derivation of the graph
	$\triangle(1,2,3) \triangle(2,4,5) \triangle(4,8,9) \triangle(5,9,10) 
	\triangle(3,6,7) \triangle(6,10,11) \triangle(7,11,12)$ in diagram notation.
    Nodes are drawn as circles with their names inscribed, nonterminal edges as boxes or triangles 
    around their 
    label, with lines to their attached nodes, and terminal edges as 
		triangles visiting their attached nodes counter-clockwise, starting at the top corner. The numbers
		inside the terminal triangles are used later to refer to the corresponding terminal edges. }
  \label{f:SP}
\end{figure}

\begin{example}[A HR Grammar for Sierpinski Triangles]
	The following rules
	$$
	\startLit \too_0 \Dlit{xyz}  \qquad 
	\Dlit{xyz} \too_1 \Dlit{xuw} \, \Dlit{uyv} \, \Dlit{wvz} \qquad
	\Dlit{xyz} \too_2 \tlit{xyz} 
	$$
	generate Sierpinski triangles as graphs where triangles are represented by ternary edges of type $\lb t$. This grammar is in fact a slightly modified version of \cite[p.~189]{habel:92} where edges of triangles are represented by binary edges.
	
	\figref{f:SP} shows a derivation with graphs as diagrams, in particular with $\lb t$-edges drawn as triangles. This corresponds to the following derivation. Underlines indicate rewritten nonterminal edges:
	\begin{align*}
		\underline{\startLit}
		& \Too_0 \underline{\nt(1,8,12)} 
		\Too_1 \nt(1,2,3) \underline{\nt(2,8,10)} \nt(3,10,12)
		\Too_1 \nt(1,2,3) \nt(2,4,5) \nt(4,8,9) \nt(5,9,10) \underline{\nt(3,10,12)} \\
		& \Too_1 \nt(1,2,3) \nt(2,4,5) \nt(4,8,9) \nt(5,9,10) \nt(3,6,7) \nt(6, 10, 11) \nt(7, 11, 12)
		 \Too_2^7 \triangle(1,2,3) \triangle(2,4,5) \triangle(4,8,9) \triangle(5,9,10) 
		        \triangle(3,6,7) \triangle(6,10,11) \triangle(7,11,12)
	\end{align*}
\end{example}

% !TEX root = main.tex

\section{Predictive Shift-Reduce Parsing}\label{s:psr}

\def\pa{p}
\def\pb{q}
\def\pc{u}
\def\pd{v}

The article \cite{drewes-hoffmann-minas-18} gives detailed definitions
and correctness proofs for PSR parsing.  Here we recall the concepts
only so far that we can describe its generalization in the next
section.

A PSR parser attempts to construct a derivation by reading
the
edges of a given input graph one after the other.%
\footnote{We silently assume that input graphs do not have isolated
  nodes. This is no real restriction as one can add special edges to
  such nodes.} %
However, the parser must not assume that the edges of the input graph
come in the same order as in a derivation.
E.g., when constructing the derivation in \figref{f:SP}, it must also
accept an input graph
$
	\triangle(1, 2, 3)
	\triangle(2, 4, 5)
	\triangle(3, 6, 7)
	\triangle(4, 8, 9)
	\triangle(5, 9, 10)
	\triangle(6, 10, 11)
	\triangle(7, 11, 12)
	$
where the edges are permuted.

Before parsing starts, a procedure described
in~\cite[Sect.~4]{Drewes-Hoffmann-Minas:16} analyzes the grammar for
the \emph{unique start node} property, by computing the possible
incidences of all nodes created by a grammar.  The unique start nodes
have to be matched by some nodes in the right-hand side of the start rule of the grammar, thus
determining where parsing begins.
For our example, the procedure detects that every Sierpinski graph has
a unique topmost node. That is a node with a single $\lb t$-edge
attached where the node is the first in the edge's attachments. The
node $x$ in the start rule $\startLit \to \Dlit{xyz}$ must be bound to
the topmost node of any input graph.%
\footnote{%
The other two nodes of the start rule, in fact, can be uniquely
identified, too, which could be used as a second and a third start node
bound to~$y$ and~$z$, respectively. However, the corresponding CFA 
is too complicated for a presentation in this paper.}
If the input graph has no topmost node, or more than one, it cannot be
a Sierpinski graph, so that parsing fails immediately.

A PSR parser is a push-down automaton that is controlled by a
\emph{characteristic finite automaton} (CFA). The stack of the PSR
parser consists of states of the CFA.  The parser makes sure that the
sequence of states on its stack always describes a valid walk through
its CFA. 

% !TEX root = main.tex

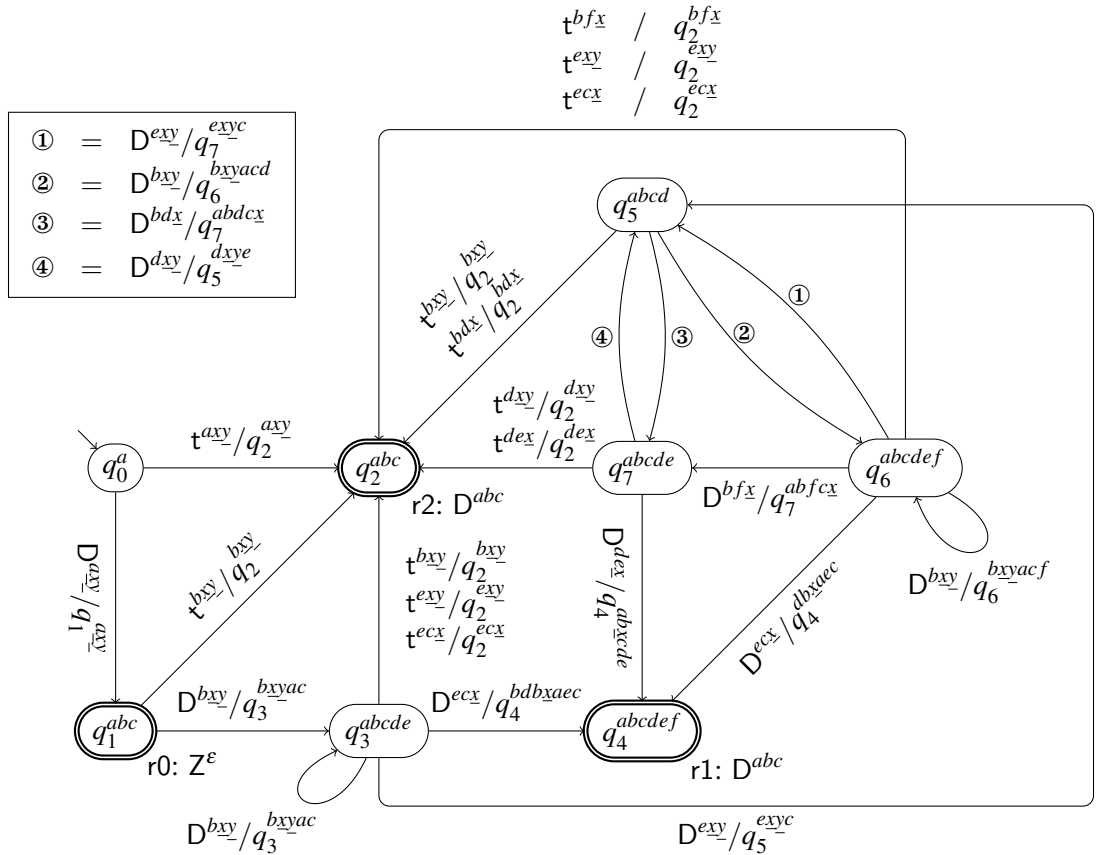
\begin{figure}[t]
  \def\x{\underline{x}}
  \def\y{\underline{y}}
	\centering
\begin{tikzpicture}[node distance=3.5cm, every node/.style={rounded rectangle},auto, ->]
  \node [draw] (q0) {$q_0^a$};
  \node [draw, thick, double, right of=q0] (q2) {$q_2^{abc}$};
  \node [draw, right of=q2] (q7) {$q_7^{abcde}$};
  \node [draw, right of=q7] (q6) {$q_6^{abcdef}$};
  \node [draw, thick, double, below of=q0] (q1) {$q_1^{abc}$};
  \node [draw, right of=q1] (q3) {$q_3^{abcde}$};
  \node [draw, thick, double, right of=q3] (q4) {$q_4^{abcdef}$};
  \node [draw, above of=q7] (q5) {$q_5^{abcd}$};
  \node [below right=1mm of q1, anchor=west] {\textsf{r0}: $\lb Z^\emptyseq$};
  \node [below right=1mm of q2, anchor=west] {\textsf{r2}: $\lb D^{abc}$};
  \node [below right=1mm of q4, anchor=west] {\textsf{r1}: $\lb D^{abc}$};

  \draw (q0) -- node[sloped,below]{$\lb D^{a\x\y} / q_1^{a\x\y}$} (q1);
  \draw (q0) -- node{$\lb t^{a\x\y} / q_2^{a\x\y}$} (q2);
  \draw (q1) -- node[sloped,above]{$\lb t^{b\x\y} / q_2^{b\x\y}$} (q2);
  \draw (q1) -- node{$\lb D^{b\x\y} / q_3^{b\x \y ac}$} (q3);
  \draw (q3) -- node[right=-4mm]{$
                \begin{array}{l@{{} / {}}l}
                  \lb t^{b\x\y} & q_2^{b\x\y}\\
                  \lb t^{e\x\y} & q_2^{e\x\y}\\
                  \lb t^{ec\x} & q_2^{ec\x}
                \end{array}
              $} (q2);
  \draw (q3) -- node{$\lb D^{ec\x} / q_4^{bdb\x aec}$} (q4);
  \draw (q5) -- node[sloped,above]{$
                \begin{array}{l@{{} / {}}l}
                  \lb t^{b\x\y} & q_2^{b\x\y}\\
                  \lb t^{bd\x} & q_2^{bd\x}
                \end{array}
              $}  (q2);
  \draw (q5) to[out=-60,in=150] node[above]{\circled 2} (q6);
  \draw (q5) to[out=-75,in=75] node[right=-1mm]{\circled 3} (q7);
  \draw (q6) to[out=120,in=-30] node[above]{\circled 1} (q5);
  \draw (q6) -- node{$\lb D^{bf\x} / q_7^{abfc\x}$} (q7);
  \draw (q6) -- node[sloped,below]{$\lb D^{ec\x} / q_4^{db\x aec}$} (q4);
  \draw (q7) -- node[above right=-1mm]{$
                \begin{array}{l@{{} / {}}l}
                  \lb t^{d\x\y} & q_2^{d\x\y}\\
                  \lb t^{de\x} & q_2^{de\x}
                \end{array}
              $} (q2);
  \draw (q7) -- node[sloped,below]{$\lb D^{de\x} / q_4^{ab\x cde}$} (q4);
  \draw (q7) to[out=105,in=-105] node[left=-1mm]{\circled 4} (q5);
  \draw[rounded corners] (q6) -- ($(q6)+(0,45mm)$) 
                              -- node[above]{$
                              \begin{array}{lcl}&&\\&&\\
                                \lb t^{bf\x} &/& q_2^{bf\x}\\
                                \lb t^{e\x\y} &/& q_2^{e\x\y}\\
                                \lb t^{ec\x} &/& q_2^{ec\x}
                              \end{array}
                              $} ($(q2)+(0,45mm)$)
                              -- (q2);
  \draw[rounded corners] (q3) -- ($(q3)+(0mm,-10mm)$) 
                              -- node[below]{$\lb D^{e\x\y} / q_5^{e\x \y c}$} ($(q3)+(95mm,-10mm)$)
                              %-- ($(q3)+(80mm,60mm)$)
                              |- (q5);
  \draw ($(q0)+(-5mm,5mm)$) -- (q0);
  \draw (q3) to [in=205,out=245,loop] node[below left]{$\lb D^{b\x\y} / q_3^{b\x \y ac}$} (q3);
  \draw (q6) to [in=290,out=330,loop] node[below]{$\lb D^{b\x\y} / q_6^{b\x \y acf}$} (q6);

  \node [left=40mm of q5,draw,rectangle] {$
  \begin{array}{lcl}
    \text{\circled 1} & = & \lb D^{e\x\y} / q_7^{e\x \y c} \\
    \text{\circled 2} & = & \lb D^{b\x\y} / q_6^{b\x \y acd}\\
    \text{\circled 3} & = & \lb D^{bd\x} / q_7^{abdc\x} \\
    \text{\circled 4} & = & \lb D^{d\x\y} / q_5^{d\x \y e} \\
    \end{array}
  $};
\end{tikzpicture}
  \caption{The characteristic finite automaton for the HR grammar of Sierpinski triangles.}
  \label{fig:auto}
\end{figure}

\figref{fig:auto} shows the CFA for our example of Sierpinski graphs.
It has been generated by the graph parser distiller
\textsc{Grappa}$^{\text{\ref{fn:grappa}}}\!\!\!,$\: using the
constructions described in~\cite{drewes-hoffmann-minas-18}, and consists
of eight states. Each state has a unique state number and a number of
\emph{parameters}, which are written as subscript and superscript,
respectively. Parameters are placeholders for nodes of the input graph,
which have already been read by the parser. The initial state is
$q_0^a$. Its parameter~$a$ is bound to the start node of the input graph, 
i.e., the topmost node, when parsing starts. Transitions between states
are labeled by pairs with a slash as a separator. The first part of a
label is the \emph{trigger} of the transition whereas the second part
of a label determines the parameters of the target state of the
transition. Note that the latter is in fact the target state of the
transition with its parameters set to the values used in the label. The
trigger is a placeholder for an edge whose attached nodes are either
parameters of the source state of the transition, or placeholders 
$\underline{x}$ or $\underline{y}$, which stand for nodes of the
input graph that have not yet been read by the parser. Note that some
transitions have multiple labels. This is in fact a shortcut for
different transitions, each with one of these labels. We are going to
describe the meaning of labels in the following and shall use
$
\triangle(1, 2, 3)
\triangle(2, 4, 5)
\triangle(3, 6, 7)
\triangle(4, 8, 9)
\triangle(5, 9, 10)
\triangle(6, 10, 11)
\triangle(7, 11, 12)
$ as an input graph.

A PSR parser starts with a stack that contains the input state with its
parameters bound to the start nodes. In our example this is $q_0^a$ with 
$a$ being bound to node~$\node{1}$, written as~$q_0^{\node{1}}$. We call 
such a state with all its parameters being
bound to input graph nodes a \emph{concrete state}. The next action of
the parser is always determined by the topmost state on the stack,
which is concrete, and by consulting the corresponding state in the
CFA. Three different types of actions are distinguished:

 A \emph{\bfseries shift} action reads a yet unread edge of the input graph.
 This corresponds to an outgoing transition with a terminal trigger. The
 trigger fits if the input graph contains an unread edge labeled with
 the trigger label and being attached to input graph nodes as specified
 by the node placeholders of the trigger. If the topmost state is
 $q_0^{\node{1}}$, there is an outgoing transition to state $q_2^{abc}$
 with a trigger $\lb t^{a\underline{x}\underline{y}}$. Parameter~$a$ is
 bound to~$\node{1}$, and its second and third attached nodes must be
 unread nodes, indicated by $\underline{x}$ and $\underline{y}$. Edge
 $\triangle(1,2,3)$ fits this trigger because $\triangle(1,2,3)$ and
 $\node 2$ as well as $\node 3$ are yet unread. The shift action marks
 this edge as well as its attached nodes as read, and pushes the
 target state of the transition on the stack. The second part of the
 label determines the binding of this state. In our example, this is
 $q_2^{a\underline{x}\underline{y}}$ where $a, \underline{x},
 \underline{y}$ are bound to $\node 1, \node 2, \node 3$, respectively.
 As a consequence, the stack will now contain $q_0^{\node{1}}$ and
 $q_2^{\node 1\node 2\node 3}$ with the latter being the new topmost
 state.

 A \emph{\bfseries reduce} action is performed when the top state of the
 stack corresponds to the right-hand side of a rule which is then
 replaced by the corresponding left-hand side. The parser recognizes
 this situation by inspecting just the topmost state of the stack; states
 that allow a reduce action are marked accordingly. In
 \figref{fig:auto}, these states are drawn with a thick border and
 additionally labeled by \textsf{r0}, \textsf{r1}, and \textsf{r2} together with
 a placeholder for a nonterminal edge. For instance, $q_2^{abc}$ is
 labeled by \textsf{r2}:$\lb D^{abc}$ where \textsf{r2} means a
 reduction using rule~2 of the grammar. The reduce action in fact
 consists of three consecutive steps. In the first step, the parser
 creates a nonterminal as indicated by the state label. In our example,
 it is \textsf{r2}:$\lb D^{abc}$. With a topmost state $q_2^{\node
 1\node 2\node 3}$, $a,b,c$ are bound to $\node 1,\node 2,\node 3$,
 which produces a nonterminal $\lb D^{\node 1\node 2\node 3}$. In the
 second step of the reduce action, the parser pops as many states off
 the stack as this rule's right-hand side contains edges, i.e., just
 one state for rule~2. For instance, when starting with stack contents 
 $q_0^{\node{1}}
 q_2^{\node 1\node 2\node 3}$, $q_2^{\node 1\node 2\node 3}$ is popped off,
 yielding a stack just containing
 $q_0^{\node{1}}$. The third step is called a \emph{goto step}. It inspects
 the new topmost state, i.e., $q_0^{\node{1}}$ here, and selects an
 outgoing transition whose trigger fits the nonterminal edge produced
 in the first step, i.e., $\lb D^{\node 1\node 2\node 3}$ and the
 transition to $q_1^{abc}$. The parser then pushes the target state
 with its parameters bound according to the transition label. In our
 example, the stack is then $q_0^{\node{1}} q_1^{\node 1\node 2\node
 3}$.
 
 An \emph{\bfseries accept} action is in fact a particular reduce action
 for the start rule, i.e., rule~0 in our example. The input graph is
 accepted if the topmost state of the stack is labeled with
 \textsf{r0}:$\lb Z^\emptyseq$, i.e., state $q_1^{abc}$ in our example, and if
 all nodes and edges of the input graph are marked as read. In our
 example with stack contents $q_0^{\node{1}} q_1^{\node 1\node
 2\node 3}$, the parser has rather reached the accepting state, but
 there are some unread edges and nodes, i.e., the input graph cannot be
 accepted yet.

The parser fails if neither a shift, reduce, nor accept
action is possible.

As described in \cite{drewes-hoffmann-minas-18}, such a CFA can be
computed for every HR grammar.\footnote{\cite{drewes-hoffmann-minas-18}
describes a simplified algorithm for computing the CFA, which may fail
to terminate for some HR grammars. \emph{Grappa}, however, employs a
more sophisticated algorithm, which can handle these grammars, too.}
But it can control a PSR parser as
described above only if its states do not have \emph{conflicts}. A
conflict is a situation where the parser must choose between different
actions. It is clear that the parser cannot run into a dead end if no
state of the CFA has a conflict; the parser can then always predict
the correct action which avoids a dead end for valid graphs.%
\footnote{This does not necessarily mean that PSR parsers are
deterministic; different edges may be chosen for the same shift action.
This does not occur in our example of Sierpinski graphs. In general, a
grammar can only be PSR parseable if it additionally satisfies the
\emph{free edge choice} property~\cite{drewes-hoffmann-minas-18}.}
But in the case of conflicts, the parser must choose between several
actions; it cannot predict the correct next action. A grammar with such
a CFA is not PSR parseable.

Our example grammar for Sierpinski graphs is not PSR parseable because
states $q_3^{abcde}$ and $q_6^{abcdef}$ have conflicts. When the parser
reaches $q_3^{abcde}$, for instance, it must read a $\lb t$-edge in the
next shift step, and it must choose between an edge being attached
to~$b$ (or rather the node that $b$ is bound to) or~$e$, indicated by
the transition to $q_2^{abc}$.

% !TEX root = main.tex

\section{Generalized Predictive Shift-Reduce Parsing}\label{s:gpsr}

In~\cite{hoffmann-minas:19} we have proposed \emph{generalized PSR}
(GPSR) parsing for grammars that are not PSR parseable. A GPSR parser
is primarily a PSR parser that follows all different choices if a state
has conflicts. It tries to save time and space in a similar way as
Tomita-style GLR parsers for context-free string grammars. Let us
briefly summarize how GPSR parsing works.

Whereas a PSR parser maintains a single stack for parsing, a GPSR
parser in fact maintains a set of stacks. This set is stored as a
so-called \emph{graph-structured stack} (GSS), which is described in the
next paragraph. For each stack, the parser determines all possible
actions based on the CFA as described for the PSR parser. The parser
has found a successful parse if the action is \emph{accept} and the
entire input graph has been read. (It may proceed if further parses
shall be found.) If the parser fails for a specific stack, the parser
just discards this stack, stops if this has been the last remaining
stack, and fails altogether if it has not found a successful parse
previously. If the CFA, however, indicates more than one possible
action, the parser duplicates the stack for each of them, and performs
each action on one of the copies.

In fact, a GPSR parser does not store complete copies of stacks, but
shares their common prefixes and suffixes. The resulting structure is a
DAG known as a \emph{graph-structured stack} (GSS) as proposed by
M.~Tomita~\cite{Tomita:85}. Each node of this DAG (called \emph{GSS node}
in the following) is a state. An individual stack is represented as a
path in the GSS, from some topmost state to the unique initial state.
Working on the GSS instead of on a set of complete copies of different
stacks does not only save space, but also time: instead of repeating
the same operations on different stacks that share the same suffix, the
parser has to perform these actions only once. Furthermore, maintaining
the GSS simplifies the construction of all parse trees (the so-called
parse forest) of an ambiguous input. But we ignore this aspect in this
paper.

Remember that we represent graphs as permutations of edges. By trying
out every action offered by the CFA in each step, the GPSR parser
effectively performs an exhaustive search in the set of all
permutations of the input graph edges permitted by the CFA. This has
two immediate effects for a GPSR parser:

\begin{enumerate}
  \item 
Consider two different stacks reached by the GPSR parser. Each stack
represents a different history of choices the parser has made. In
particular, different input graph edges may have been read in these
histories. The parser, therefore, cannot globally mark edges as read, 
but it must store, for each stack separately, which
edges of the input graph have been read. In fact, each GSS node keeps
track of the set of input graph edges that have been read so far. Note
that GSS nodes may be shared only if both their concrete states and
their sets of read edges coincide.
  
\item Whenever the parser has a GSS that represents at least two
different stacks, it must choose the stack that it considers next for
its actions. It may, for instance, employ a breadth-first strategy or a
depth-first strategy. This is in fact the major difference between GLR
parsers for context-free string grammars and GPSR parsers for HR
grammars: Whenever a GLR parser executes a shift action, this is done
for all top-level GSS nodes ``simultaneously''. And it then performs
all possible reduce actions before the next shift action is executed.
As a consequence, each stack encoded in the GSS represents a parse of
the same substring of the input string. This is not the case for GPSR
parsers. They may be rather forced to try out several reading sequences
of the input graph, which may result in exponential complexity.

In~\cite{hoffmann-minas:19}, we have shown for two example
languages (series-parallel graphs and structured flowcharts; see
\sectref{s:eval}) that the chosen strategy strongly affects the parser
speed. In fact, a standard strategy was always too slow, even slower
than a simple CYK parser.
Instead, specifically tailored strategies have been used that give
certain grammar rules preference over others. This requires extra
manual work when building a parser and was the motivation for this
paper, in particular because even this does not always help in creating a GPSR
parser that is faster than a CYK parser. 
\end{enumerate}

As a matter of fact, breadth-first and depth-first produce slow parsers
for the language of Sierpinski graphs, too. We shall demonstrate this by
describing the steps of the GPSR parser for the input graph
$\triangle(1, 2, 3) \triangle(2, 4, 5) \triangle(3, 6, 7) \triangle(4,
8, 9) \triangle(5, 9, 10) \triangle(6, 10, 11) \triangle(7, 11, 12)$
(see \figref{f:SP}). To save space, we refer to these edges by numbers
$1=\triangle(1, 2, 3), 2=\triangle(2, 4, 5), 3=\triangle(3, 6, 7),
4=\triangle(4, 8, 9), 5=\triangle(5, 9, 10), 6=\triangle(6, 10, 11),
7=\triangle(7, 11, 12)$. These numbers correspond to the numbers within
the triangles in \figref{f:SP}. And we write GSS nodes in compact form:
e.g., $\state{2}{\node{4}\node{8}\node{9}}{124}$ refers to the concrete
state $q_2^{\node{4}\node{8}\node{9}}$ and indicates that the edges
$1=\triangle(1, 2, 3)$, $2=\triangle(2, 4, 5)$, and $4=\triangle(4, 8,
9)$ have been read already. \figref{fig:gpsr-parse} shows the
graph-structured stacks after each step of the GPSR parser where a step
consists of all actions performed by the parser when working on a
specific state. Stacks grow to the right, i.e., the initial state is at
the left end whereas topmost states are at the right ends. The steps in
fact follow the depth-first strategy which turned out to be a bit
faster than the breadth-first strategy.

% !TEX root = main.tex

\begin{figure}[t]
\def\edge#1{#1}
\def\gss#1#2{
   #1 & 
\begin{tikzpicture}
  [
    baseline=(root.base),
    GSSNode/.style          = {font=\normalsize},
    grow                    = right,
    sibling distance        = 6mm,
    growth parent anchor={east}, 
    nodes={anchor=west},
    level distance          = 2.7mm,
    edge from parent/.style = {draw},
    inner ysep=3pt,
    inner xsep=2pt
  ]
  \input{#2}
\end{tikzpicture}
\\\hline
}
\hspace*{\fill}
\begin{tabular}[t]{|@{\;}c@{\;}|@{}l@{}|}
  \hline
  
    \footnotesize 0 & 
  \begin{tikzpicture}
    [
      baseline=(root.base),
      GSSNode/.style          = {font=\footnotesize},
      grow                    = right,
      sibling distance        = 5mm,
      growth parent anchor={east}, 
      nodes={anchor=west},
      level distance          = 1.9mm,
      edge from parent/.style = {draw},
      inner ysep=1pt,
      inner xsep=1pt
    ]
    \input{gss/gpsr/sierpinski_gpsr-gss-00}
  \end{tikzpicture}
  \\\hline

    \footnotesize 1 & 
  \begin{tikzpicture}
    [
      baseline=(root.base),
      GSSNode/.style          = {font=\footnotesize},
      grow                    = right,
      sibling distance        = 5mm,
      growth parent anchor={east}, 
      nodes={anchor=west},
      level distance          = 1.9mm,
      edge from parent/.style = {draw},
      inner ysep=1pt,
      inner xsep=1pt
    ]
    \input{gss/gpsr/sierpinski_gpsr-gss-01}
  \end{tikzpicture}
  \\\hline

    \footnotesize 2 & 
  \begin{tikzpicture}
    [
      baseline=(root.base),
      GSSNode/.style          = {font=\footnotesize},
      grow                    = right,
      sibling distance        = 5mm,
      growth parent anchor={east}, 
      nodes={anchor=west},
      level distance          = 1.9mm,
      edge from parent/.style = {draw},
      inner ysep=1pt,
      inner xsep=1pt
    ]
    \input{gss/gpsr/sierpinski_gpsr-gss-02}
  \end{tikzpicture}
  \\\hline

    \footnotesize 3 & 
  \begin{tikzpicture}
    [
      baseline=(root.base),
      GSSNode/.style          = {font=\footnotesize},
      grow                    = right,
      sibling distance        = 5mm,
      growth parent anchor={east}, 
      nodes={anchor=west},
      level distance          = 1.9mm,
      edge from parent/.style = {draw},
      inner ysep=1pt,
      inner xsep=1pt
    ]
    \input{gss/gpsr/sierpinski_gpsr-gss-03}
  \end{tikzpicture}
  \\\hline

    \footnotesize 4 & 
  \begin{tikzpicture}
    [
      baseline=(root.base),
      GSSNode/.style          = {font=\footnotesize},
      grow                    = right,
      sibling distance        = 5mm,
      growth parent anchor={east}, 
      nodes={anchor=west},
      level distance          = 1.9mm,
      edge from parent/.style = {draw},
      inner ysep=1pt,
      inner xsep=1pt
    ]
    \input{gss/gpsr/sierpinski_gpsr-gss-04}
  \end{tikzpicture}
  \\\hline

    \footnotesize 5 & 
  \begin{tikzpicture}
    [
      baseline=(root.base),
      GSSNode/.style          = {font=\footnotesize},
      grow                    = right,
      sibling distance        = 5mm,
      growth parent anchor={east}, 
      nodes={anchor=west},
      level distance          = 1.9mm,
      edge from parent/.style = {draw},
      inner ysep=1pt,
      inner xsep=1pt
    ]
    \input{gss/gpsr/sierpinski_gpsr-gss-05}
  \end{tikzpicture}
  \\\hline

    \footnotesize 6 & 
  \begin{tikzpicture}
    [
      baseline=(root.base),
      GSSNode/.style          = {font=\footnotesize},
      grow                    = right,
      sibling distance        = 5mm,
      growth parent anchor={east}, 
      nodes={anchor=west},
      level distance          = 1.9mm,
      edge from parent/.style = {draw},
      inner ysep=1pt,
      inner xsep=1pt
    ]
    \input{gss/gpsr/sierpinski_gpsr-gss-06}
  \end{tikzpicture}
  \\\hline

    \footnotesize 7 & 
  \begin{tikzpicture}
    [
      baseline=(root.base),
      GSSNode/.style          = {font=\footnotesize},
      grow                    = right,
      sibling distance        = 5mm,
      growth parent anchor={east}, 
      nodes={anchor=west},
      level distance          = 1.9mm,
      edge from parent/.style = {draw},
      inner ysep=1pt,
      inner xsep=1pt
    ]
    \input{gss/gpsr/sierpinski_gpsr-gss-07}
  \end{tikzpicture}
  \\\hline

    \footnotesize 8 & 
  \begin{tikzpicture}
    [
      baseline=(root.base),
      GSSNode/.style          = {font=\footnotesize},
      grow                    = right,
      sibling distance        = 5mm,
      growth parent anchor={east}, 
      nodes={anchor=west},
      level distance          = 1.9mm,
      edge from parent/.style = {draw},
      inner ysep=1pt,
      inner xsep=1pt
    ]
    \input{gss/gpsr/sierpinski_gpsr-gss-08}
  \end{tikzpicture}
  \\\hline

    \footnotesize 9 & 
  \begin{tikzpicture}
    [
      baseline=(root.base),
      GSSNode/.style          = {font=\footnotesize},
      grow                    = right,
      sibling distance        = 5mm,
      growth parent anchor={east}, 
      nodes={anchor=west},
      level distance          = 1.9mm,
      edge from parent/.style = {draw},
      inner ysep=1pt,
      inner xsep=1pt
    ]
    \input{gss/gpsr/sierpinski_gpsr-gss-09}
  \end{tikzpicture}
  \\\hline

    \footnotesize 10 & 
  \begin{tikzpicture}
    [
      baseline=(root.base),
      GSSNode/.style          = {font=\footnotesize},
      grow                    = right,
      sibling distance        = 5mm,
      growth parent anchor={east}, 
      nodes={anchor=west},
      level distance          = 1.9mm,
      edge from parent/.style = {draw},
      inner ysep=1pt,
      inner xsep=1pt
    ]
    \input{gss/gpsr/sierpinski_gpsr-gss-10}
  \end{tikzpicture}
  \\\hline
  
\end{tabular}
\hspace*{\fill}
\begin{tabular}[t]{|@{\;}c@{\;}|@{}l@{}|}
  \hline
  
    \footnotesize 11 & 
  \begin{tikzpicture}
    [
      baseline=(root.base),
      GSSNode/.style          = {font=\footnotesize},
      grow                    = right,
      sibling distance        = 5mm,
      growth parent anchor={east}, 
      nodes={anchor=west},
      level distance          = 1.9mm,
      edge from parent/.style = {draw},
      inner ysep=1pt,
      inner xsep=1pt
    ]
    \input{gss/gpsr/sierpinski_gpsr-gss-11}
  \end{tikzpicture}
  \\\hline

    \footnotesize 12 & 
  \begin{tikzpicture}
    [
      baseline=(root.base),
      GSSNode/.style          = {font=\footnotesize},
      grow                    = right,
      sibling distance        = 5mm,
      growth parent anchor={east}, 
      nodes={anchor=west},
      level distance          = 1.9mm,
      edge from parent/.style = {draw},
      inner ysep=1pt,
      inner xsep=1pt
    ]
    \input{gss/gpsr/sierpinski_gpsr-gss-12}
  \end{tikzpicture}
  \\\hline

    \footnotesize 13 & 
  \begin{tikzpicture}
    [
      baseline=(root.base),
      GSSNode/.style          = {font=\footnotesize},
      grow                    = right,
      sibling distance        = 5mm,
      growth parent anchor={east}, 
      nodes={anchor=west},
      level distance          = 1.9mm,
      edge from parent/.style = {draw},
      inner ysep=1pt,
      inner xsep=1pt
    ]
    \input{gss/gpsr/sierpinski_gpsr-gss-13}
  \end{tikzpicture}
  \\\hline

    \footnotesize 14 & 
  \begin{tikzpicture}
    [
      baseline=(root.base),
      GSSNode/.style          = {font=\footnotesize},
      grow                    = right,
      sibling distance        = 5mm,
      growth parent anchor={east}, 
      nodes={anchor=west},
      level distance          = 1.9mm,
      edge from parent/.style = {draw},
      inner ysep=1pt,
      inner xsep=1pt
    ]
    \input{gss/gpsr/sierpinski_gpsr-gss-14}
  \end{tikzpicture}
  \\\hline

    \footnotesize 15 & 
  \begin{tikzpicture}
    [
      baseline=(root.base),
      GSSNode/.style          = {font=\footnotesize},
      grow                    = right,
      sibling distance        = 5mm,
      growth parent anchor={east}, 
      nodes={anchor=west},
      level distance          = 1.9mm,
      edge from parent/.style = {draw},
      inner ysep=1pt,
      inner xsep=1pt
    ]
    \input{gss/gpsr/sierpinski_gpsr-gss-15}
  \end{tikzpicture}
  \\\hline

    \footnotesize 16 & 
  \begin{tikzpicture}
    [
      baseline=(root.base),
      GSSNode/.style          = {font=\footnotesize},
      grow                    = right,
      sibling distance        = 5mm,
      growth parent anchor={east}, 
      nodes={anchor=west},
      level distance          = 1.9mm,
      edge from parent/.style = {draw},
      inner ysep=1pt,
      inner xsep=1pt
    ]
    \input{gss/gpsr/sierpinski_gpsr-gss-16}
  \end{tikzpicture}
  \\\hline

    \footnotesize 17 & 
  \begin{tikzpicture}
    [
      baseline=(root.base),
      GSSNode/.style          = {font=\footnotesize},
      grow                    = right,
      sibling distance        = 5mm,
      growth parent anchor={east}, 
      nodes={anchor=west},
      level distance          = 1.9mm,
      edge from parent/.style = {draw},
      inner ysep=1pt,
      inner xsep=1pt
    ]
    \input{gss/gpsr/sierpinski_gpsr-gss-17}
  \end{tikzpicture}
  \\\hline

    \footnotesize 18 & 
  \begin{tikzpicture}
    [
      baseline=(root.base),
      GSSNode/.style          = {font=\footnotesize},
      grow                    = right,
      sibling distance        = 5mm,
      growth parent anchor={east}, 
      nodes={anchor=west},
      level distance          = 1.9mm,
      edge from parent/.style = {draw},
      inner ysep=1pt,
      inner xsep=1pt
    ]
    \input{gss/gpsr/sierpinski_gpsr-gss-18}
  \end{tikzpicture}
  \\\hline

    \footnotesize 19 & 
  \begin{tikzpicture}
    [
      baseline=(root.base),
      GSSNode/.style          = {font=\footnotesize},
      grow                    = right,
      sibling distance        = 5mm,
      growth parent anchor={east}, 
      nodes={anchor=west},
      level distance          = 1.9mm,
      edge from parent/.style = {draw},
      inner ysep=1pt,
      inner xsep=1pt
    ]
    \input{gss/gpsr/sierpinski_gpsr-gss-19}
  \end{tikzpicture}
  \\\hline

    \footnotesize 20 & 
  \begin{tikzpicture}
    [
      baseline=(root.base),
      GSSNode/.style          = {font=\footnotesize},
      grow                    = right,
      sibling distance        = 5mm,
      growth parent anchor={east}, 
      nodes={anchor=west},
      level distance          = 1.9mm,
      edge from parent/.style = {draw},
      inner ysep=1pt,
      inner xsep=1pt
    ]
    \input{gss/gpsr/sierpinski_gpsr-gss-20}
  \end{tikzpicture}
  \\\hline

    \footnotesize 21 & 
  \begin{tikzpicture}
    [
      baseline=(root.base),
      GSSNode/.style          = {font=\footnotesize},
      grow                    = right,
      sibling distance        = 5mm,
      growth parent anchor={east}, 
      nodes={anchor=west},
      level distance          = 1.9mm,
      edge from parent/.style = {draw},
      inner ysep=1pt,
      inner xsep=1pt
    ]
    \input{gss/gpsr/sierpinski_gpsr-gss-21}
  \end{tikzpicture}
  \\\hline

    \footnotesize 22 & 
  \begin{tikzpicture}
    [
      baseline=(root.base),
      GSSNode/.style          = {font=\footnotesize},
      grow                    = right,
      sibling distance        = 5mm,
      growth parent anchor={east}, 
      nodes={anchor=west},
      level distance          = 1.9mm,
      edge from parent/.style = {draw},
      inner ysep=1pt,
      inner xsep=1pt
    ]
    \input{gss/gpsr/sierpinski_gpsr-gss-22}
  \end{tikzpicture}
  \\\hline

    \footnotesize 23 & 
  \begin{tikzpicture}
    [
      baseline=(root.base),
      GSSNode/.style          = {font=\footnotesize},
      grow                    = right,
      sibling distance        = 5mm,
      growth parent anchor={east}, 
      nodes={anchor=west},
      level distance          = 1.9mm,
      edge from parent/.style = {draw},
      inner ysep=1pt,
      inner xsep=1pt
    ]
    \input{gss/gpsr/sierpinski_gpsr-gss-23}
  \end{tikzpicture}
  \\\hline

    \footnotesize 24 & 
  \begin{tikzpicture}
    [
      baseline=(root.base),
      GSSNode/.style          = {font=\footnotesize},
      grow                    = right,
      sibling distance        = 5mm,
      growth parent anchor={east}, 
      nodes={anchor=west},
      level distance          = 1.9mm,
      edge from parent/.style = {draw},
      inner ysep=1pt,
      inner xsep=1pt
    ]
    \input{gss/gpsr/sierpinski_gpsr-gss-24}
  \end{tikzpicture}
  \\\hline
  
\end{tabular}
\hspace*{\fill}
\caption{Graph-structured stacks and steps of the GPSR parser when parsing the Sierpinski graph with the edges $1=\triangle(1, 2,
3), 2=\triangle(2, 4, 5), 3=\triangle(3, 6, 7), 4=\triangle(4, 8, 9),
5=\triangle(5, 9, 10), 6=\triangle(6, 10, 11), 7=\triangle(7, 11, 12)$.}
\label{fig:gpsr-parse}
\end{figure}

The parser starts (step~0) with a single stack that contains just
$\state{0}{\node{1}}{\emptyset}$, i.e., the initial (concrete) state
$q_0^{\node 1}$ where no edge has been read yet. The first four steps
are just PSR steps as described in the previous section: edge
$1=\triangle(1, 2, 3)$ is shifted in step~1, a reduce action for rule~2
happens in step~2. Edge $2=\triangle(2, 4, 5)$ is shifted in step~3,
and this edge is reduced using rule~2 in step~4, reaching state
$q_3^{\node{2}\node{4}\node{5}\node{1}\node{3}}$. This state allows to
shift $3=\triangle(3, 6, 7)$ as well as $4=\triangle(4, 8, 9)$ (see
\figref{fig:auto}), producing two stacks, represented by the GSS after
step~5. Note that the topmost states of these stacks are both
$q_2$-states, but with different parameter bindings and differing sets
of read input graph edges. $\state{2}{\node{3}\node{6}\node{7}}{123}$
is reduced in step~6, resulting in
$\state{5}{\node{3}\node{6}\node{7}\node{5}}{123}$, which is considered
next in step 7 because of the depth-first strategy. In fact, the parser
continues working on this stack until it fails in step~12: the stack
has the topmost state
$\state{5}{\node{3}\node{10}\node{12}\node{5}}{12367}$ when step~12 starts, i.e., only
$4=\triangle(4, 8, 9)$ and $5=\triangle(5, 9, 10)$ are yet unread, but
they do not fit to any outgoing transition of
$q_5^{\node{3}\node{10}\node{12}\node{5}}$. The parser continues 
working on the remaining stack, i.e., with topmost state
$\state{2}{\node{4}\node{8}\node{9}}{124}$. The remaining steps are
again plain PSR steps because the parser does not need to choose between
different actions until it accepts the input graph in step~24 in
state~$\state{1}{\node{1}\node{8}\node{12}}{1234567}$, i.e., the
accepting state with all edges having been read.

The parser has in fact wasted time by choosing the topmost state
$\state{2}{\node{3}\node{6}\node{7}}{123}$ for the next stack to work
on in step~6. If it had chosen $\state{2}{\node{4}\node{8}\node{9}}{124}$
instead, it would have eventually reached the following GSS in step~17:

\begin{center}
\begin{tikzpicture}
  [
    baseline=(root.base),
    GSSNode/.style          = {font=\normalsize},
    grow                    = right,
    sibling distance        = 6mm,
    growth parent anchor={east}, 
    nodes={anchor=west},
    level distance          = 3mm,
    edge from parent/.style = {draw},
    inner ysep=3pt,
    inner xsep=2pt
  ]
\node [GSSNode] (root) {$\state{0}{\node{1}}{\emptyset}$}
   child { node [GSSNode] {$\state{1}{\node{1}\node{8}\node{12}}{1234567}$}}
   child { node [GSSNode] {$\state{1}{\node{1}\node{2}\node{3}}{1}$}
      child { node [GSSNode] {$\state{3}{\node{2}\node{4}\node{5}\node{1}\node{3}}{12}$}
         child { node [GSSNode] {$\state{2}{\node{4}\node{8}\node{9}}{124}$}}}};
\end{tikzpicture}
\end{center}

\noindent i.e., the parser would have found a successful parse 
after 17 instead of 24 steps. Of course, the parser could then continue
with the remaining stack, which would correspond to the steps~6--12 in
\figref{fig:gpsr-parse}., i.e., it would not produce further results.
The parser can terminate as soon as it has found a parse because the
grammar is unambiguous. But even if the grammar were ambiguous, the
parser could terminate after the first parse being found if one is
interested in just one parse.

\begin{figure}[t]
	\centering\includegraphics[width=0.85\textwidth]{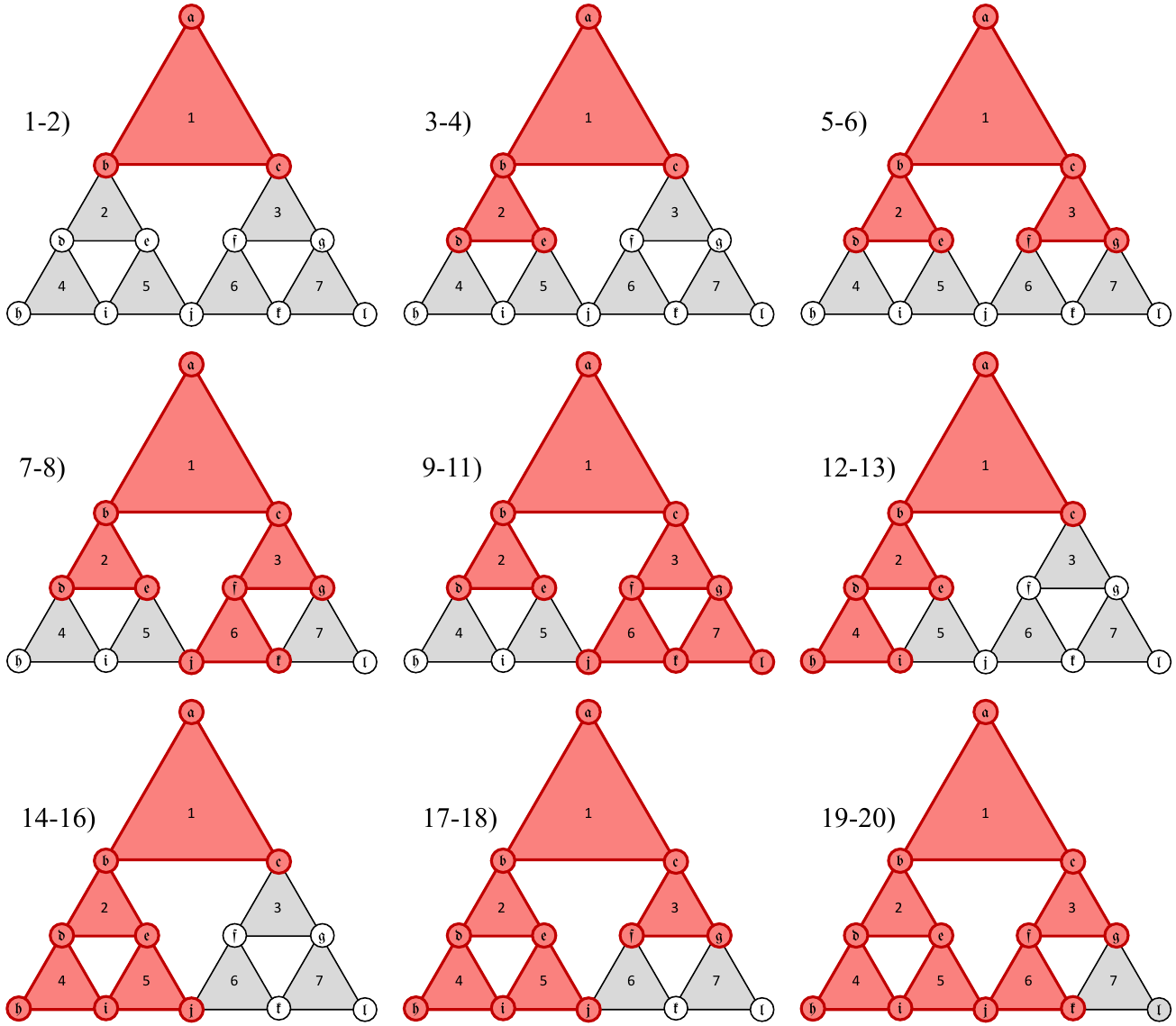} 
	\caption{Visual representation of the edges that are marked as read after the different steps in \figref{fig:gpsr-parse}.}
	\label{f:gpsr-progress}
\end{figure}

So the question remains whether the parser could be improved by more
carefully choosing the stack where the parser continues. For this
purpose, consider \figref{f:gpsr-progress}, which shows the diagram of
the input graph after steps~1--20 in \figref{fig:gpsr-parse}. It
highlights those edges that are marked as read in the state that has
just been pushed to the GSS in the corresponding step. In steps~1
and~2, for instance, it is the topmost triangle $1=\triangle(1, 2, 3)$,
in steps~3 and~4 triangles $1=\triangle(1, 2, 3)$ as well as
$2=\triangle(2, 4, 5)$, and so on. \figref{f:gpsr-progress} does not
show the situation for steps 21--24 where all edges are marked as read.
As one can see, the parser erroneously ``walks down'' to the lower right
triangles~3, 6, and~7 in steps~6--11, discards the corresponding stack
in step~12, and again walks down the same path in steps~17--20. So one
could assume that a strategy that chooses the left walk first (which
corresponds to state~$\state{2}{\node{4}\node{8}\node{9}}{124}$)
\begin{figure}[t]
  \centering%
  \includegraphics[width=4cm]{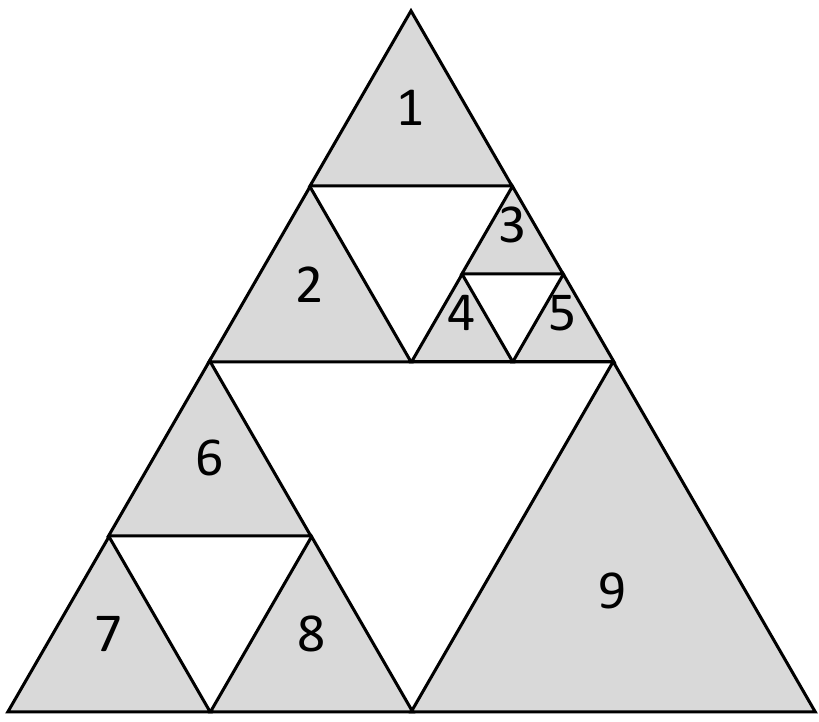}%
  \caption{A Sierpinski graph where ``walking down right first'' is faster than ``walking down left first''.}
  \label{fig:right-faster}
\end{figure}
instead of the right walk (which corresponds to
state~$\state{2}{\node{3}\node{6}\node{7}}{123}$) would improve the
parser behavior. However, this is in general not the case. \figref{fig:right-faster} shows a Sierpinski graph where walking down right first
finds the parse faster than walking down left first (25 vs.~30 steps):
% triangles 1--5 must be reduced before reducing triangles 6--8 in a
% successful parse. But 
Walking left down first reduces triangles 6--8
twice, once before reducing triangles 3--5 and once after that, which is
avoided when walking right down first.
%because the triangles 1--5 must be reduced before triangles 6, 7, and
%8 can be handled. But this requires handling of triangles 3, 4, and 5
%before triangle~6.
Apparently, there is not an easy strategy to always find the parse
fast. But memoization solves this problem.

% !TEX root = main.tex

\section{Memoization}\label{s:memo}

The GPSR parser finds a parse for a valid input graph faster if it
either avoids dead ends like the erroneous walk right down in steps
6--11 (\figref{f:gpsr-progress}) or if the effort spent in such a dead
end is not wasted, but is reused later. To see this, let us
consider the parsing steps in \figref{fig:gpsr-parse} more closely. In step~11,
it performs a reduce action for rule~1 on state
$\state{4}{\node{3}\node{10}\node{12}\node{6}\node{7}\node{11}}{12367}$
producing a nonterminal $\lb D^{\node{3}\node{10}\node{12}}$ (see
\figref{fig:auto}), popping three states off the stack yielding
$\state{3}{\node{2}\node{4}\node{5}\node{1}\node{3}}{12}$ as a
(temporary) topmost state and then pushes state
$\state{5}{\node{3}\node{10}\node{12}\node{5}}{12367}$. Moreover, it is
known, by comparing the set of read edges of this new topmost state
with the set of its predecessor on the stack, that $\lb
D^{\node{3}\node{10}\node{12}}$ represents the subgraph consisting of
the triangles~3, 6, and~7. But the same nonterminal $\lb
D^{\node{3}\node{10}\node{12}}$ representing the same subgraph is again
produced in step~23 where the parser performs a reduce action for
rule~1 on state
$\state{4}{\node{3}\node{10}\node{12}\node{6}\node{7}\node{11}}{1234567}$,
pops three states off the stack yielding state
$\state{3}{\node{2}\node{8}\node{10}\node{1}\node{3}}{1245}$, and
performs a goto step to
$\state{4}{\node{1}\node{8}\node{12}\node{2}\node{3}\node{10}}{1234567}$
triggered by $\lb D^{\node{3}\node{10}\node{12}}$ (see
\figref{fig:auto}). Note, however, that
$\state{3}{\node{2}\node{8}\node{10}\node{1}\node{3}}{1245}$ was
already the topmost state after step~16. So if the parser remembered
that it has produced a $\lb D^{\node{3}\node{10}\node{12}}$ earlier, it
could reuse it in step~17 and perform a goto step to
$\state{4}{\node{1}\node{8}\node{12}\node{2}\node{3}\node{10}}{1234567}$
right away. As a result, the parser would immediately reach the GSS
that \figref{fig:gpsr-parse} shows after step~23, i.e., the parser
would skip six steps and accept the input graph in 18 instead of 24
steps. 

In the following, we describe how this observation leads to a
systematic approach that allows to skip entire sequences of parsing steps
by reusing nonterminal edges that have been produced earlier. This is a
\emph{memoization} approach because it depends on memorizing these
nonterminal edges.

The main idea is to store a nonterminal edge in a \emph{memo store}
whenever it is produced in a reduce action and to look up nonterminals 
in the memo store whenever the parser reaches a state with an outgoing
transition triggered by nonterminal edges. The memo store in fact must
store nonterminal edges together with the set of terminal edges that
are represented by them. To be more precise, let us assume that the
parser analyzes the input graph $h \in \Gr\TLab$. The memo store then
contains pairs $\tup{\literal A, g}$ where $\literal A$ is a
nonterminal edge, $g \in \Gr\TLab$ is a terminal graph with $\literal A
\der^* g$ and $h \perm g h'$ for some graph $h' \in \Gr\TLab$, i.e.,
$g$ consists of input graph edges. For instance, the memo store after step~16 in
\figref{fig:gpsr-parse} consists of the following pairs,
produced by the reduce actions in one of the previous steps:
\begin{align*}
	\{ &
	\tup{\lb D^{\node{2}\node{8}\node{10}}\!, 245},
	\tup{\lb D^{\node{2}\node{4}\node{5}}\!, 2},
	\tup{\lb D^{\node{3}\node{10}\node{12}}\!, 367},
	\tup{\lb D^{\node{3}\node{6}\node{7}}\!, 3},\\&
	\tup{\lb D^{\node{1}\node{2}\node{3}}\!, 1},
	\tup{\lb D^{\node{4}\node{8}\node{9}}\!, 4},
	\tup{\lb D^{\node{5}\node{9}\node{10}}\!, 5},
	\tup{\lb D^{\node{6}\node{10}\node{11}}\!, 6},
	\tup{\lb D^{\node{7}\node{11}\node{12}}\!, 7}
		\},
\end{align*}
Edges in the second components of pairs are again represented by their
numbers.

The lookup operation is controlled by the nodes bound to parameters of
the current state, by the (nonterminal) label of the transition, and by
the set~$R$ of edges that are marked as read in the current state. The
lookup operation may return \emph{valid} pairs only. These are pairs
$\tup{\literal A, g}$ whose graph~$g$ does not contain any edge that
is also a member of~$R$. Otherwise, edges in $g$ and in $R$ would be read
twice.%
\footnote{We assume that there are no parallel edges with the same
label. Otherwise, each edge must have a unique name and the lookup
operation must make sure that it does not return an edge with a name
that is also a member of~$R$.}

As an example, let us now consider state
$\state{3}{\node{2}\node{8}\node{10}\node{1}\node{3}}{1245}$ after step~16. The CFA
(\figref{fig:auto}) has three outgoing transitions with nonterminal
triggers $\lb D^{\node{8}\underline{x}\underline{y}}$, $\lb
D^{\node{3}\node{10}\underline{x}}$, and $\lb
D^{\node{3}\underline{x}\underline{y}}$ when replacing parameters by
nodes bound to them. $\underline{x}$ and $\underline{y}$ may be bound
only to nodes that have not yet been read in state
$\state{3}{\node{2}\node{8}\node{10}\node{1}\node{3}}{1245}$. Unread
nodes are determined by the set 1245 of read edges, i.e., $\node{6},
\node{7}, \node{11}, \node{12}$ are unread in this state. The memo
store, therefore, does not contain a pair for $\lb
D^{\node{8}\underline{x}\underline{y}}$, but it contains $\tup{\lb
D^{\node{3}\node{10}\node{12}}\!, 367}$ for $\lb
D^{\node{3}\node{10}\underline{x}}$ and $\tup{\lb
D^{\node{3}\node{6}\node{7}}\!, 3}$ for $\lb
D^{\node{3}\underline{x}\underline{y}}$. Note that $\tup{\lb
D^{\node{3}\node{10}\node{12}}\!, 367}$ does not fit $\lb
D^{\node{3}\underline{x}\underline{y}}$ because node~$\node{10}$ has
been read already. The lookup operation, therefore, has in fact found
two valid pairs, and the parser could perform goto actions with both of
them. Moreover, it could ignore them both and continue in the regular
way, i.e., shift edge $3=\triangle(3, 6, 7)$ (see step~17 in
\figref{fig:gpsr-parse}). Because the GPSR parser, by design, does not
rule out any choice, it will consider all of the three choices here. That way,
memoization does not affect the correctness of the parser; if reusing
of nonterminals does not lead to acceptance of a valid input graph,
regular GPSR will do. But the parser needs a criterion which of the
choices to try first. The obvious criterion is to prioritize the nonterminal
edge that represents the largest subgraph; the corresponding goto
has the potential to skip the longest sequence of parsing steps. In our
example, this is $\tup{\lb D^{\node{3}\node{10}\node{12}}\!, 367}$, i.e.,
just the case described at the beginning of this section.

The \textsc{Grappa}$^{\text{\ref{fn:grappa}}}\!\!\!$\: parser distiller has been extended to generate
parsers that maintain a memo store in hash tables and that look up all
valid pairs when the parser reaches a state with outgoing nonterminal
edges. Looked up pairs are ordered by the size of their represented subgraph
and tried in that sequence. And it tries the regular GPSR actions if
none of these choices leads to acceptance.

% !TEX root = main.tex

\section {Evaluation}
\label{s:eval}

We now report on running time experiments when parsing Sierpinski
graphs. We generated three different parsers: a
CYK parser using
\textsc{DiaGen}\footnote{Homepage:
\href{https://www.unibw.de/inf2/diagen/}{www.unibw.de/inf2/diagen}}~\cite{Minas:01a}, a
GPSR parser using the depth-first strategy described in
\sectref{s:gpsr}, and finally a GPSR parser using the depth-first
strategy and memoization as described in the previous section. The GPSR
parsers have been generated using
\textsc{Grappa}, and they stop as soon they can accept the input graph.  The CYK parser
was in fact optimized in two ways: the parser creates nonterminal edges
by dynamic programming, and each of these edges can be derived to a
certain subgraph of the input graph. The optimized parser makes sure
that it does not create two or more indistinguishable nonterminals for
the same subgraph, even if the nonterminals represent different
derivation trees (which does not occur here.) And it stops as soon as it finds a derivation
of the entire input graph.

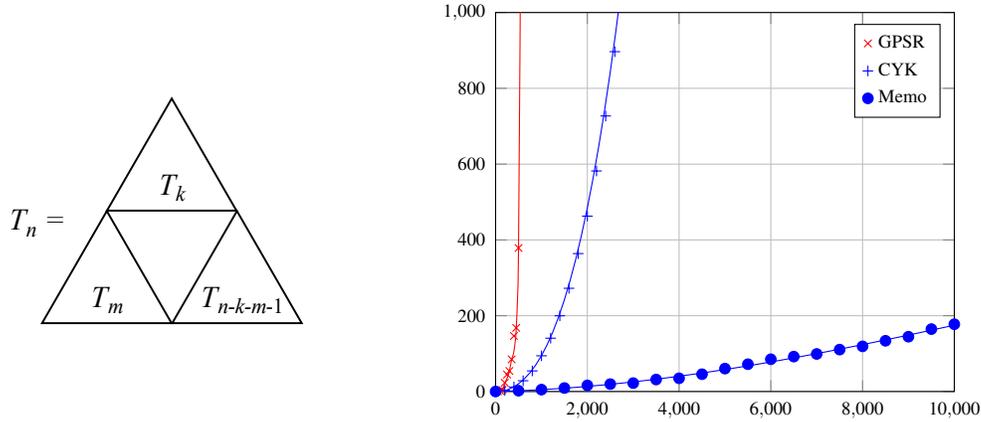
\begin{figure}[t]
	\centering{
	\begin{minipage}{0.25\textwidth}
	\includegraphics[width=\textwidth]{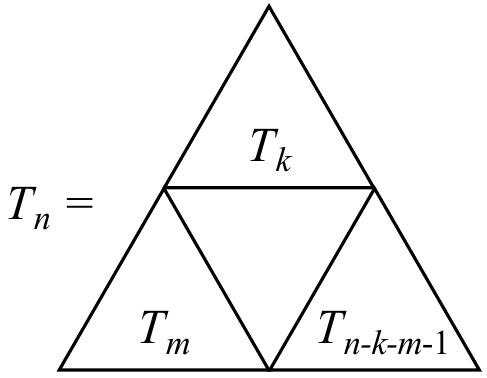}
	\end{minipage}
	\hspace{1.5cm}
	\begin{minipage}{0.48\textwidth}
% !TEX root = main.tex
%
	\begin{tikzpicture}
	\begin{axis}[
	  width=\textwidth, 
	  xmin=0, xmax=10000, 
	  ymin=0, ymax=1000, 
	  legend style={legend pos=north east}
	  ]
	% Cmd line: de.unibwm.inf2.grappa.examples.sierpinski.SierpinskiTest -graph uniform -p GPSR_LIFO -pt -s 50 -start 0 -r 20 -stat -shuffle -w -top 2 550
	\addplot [mark=x,red,only marks] coordinates {
	  (0, 0.021556)
		(50, 0.415)
		(100, 1.645444)
		(150, 3.863778)
		(200, 23.074278)
		(250, 44.742611)
		(300, 54.224722)
		(350, 84.801)
		(400, 146.520556)
		(450, 167.9285)
		(500, 378.708389)
		(550, 1288.034833)};
	\addlegendentry{GPSR}
	%  Cmd line: cyk.sierpinski.SierpinskiTest -graph uniform -s 200 -pt -r 15 -stat -start 0 -top 2 3000
	\addplot [mark=+,blue,only marks] coordinates {
	  (0, 0.02)
		(200, 2.967692)
		(400, 11.946)
		(600, 28.422923)
		(800, 54.350231)
		(1000, 94.255692)
		(1200, 140.712769)
		(1400, 199.835923)
		(1600, 272.381615)
		(1800, 363.989538)
		(2000, 462.729923)
		(2200, 581.875308)
		(2400, 727.067077)
		(2600, 896.711462)
		(2800, 1101.195231)
		(3000, 1336.664846)};
	\addlegendentry{CYK}
	% Cmd line: de.unibwm.inf2.grappa.examples.sierpinski.SierpinskiTest -graph uniform -p memo -pt -s 500 -start 0 -r 20 -stat -shuffle -w -top 2 10000
	\addplot [mark=*,blue,only marks] coordinates {
	  (0, 0.032722)
		(500, 2.115833)
		(1000, 5.009056)
		(1500, 9.214111)
		(2000, 16.160444)
		(2500, 19.548611)
		(3000, 22.196333)
		(3500, 31.631667)
		(4000, 34.8885)
		(4500, 45.668778)
		(5000, 60.423944)
		(5500, 72.005)
		(6000, 85.127111)
		(6500, 92.132278)
		(7000, 99.082833)
		(7500, 110.586778)
		(8000, 119.175167)
		(8500, 133.889833)
		(9000, 144.628278)
		(9500, 164.661222)
		(10000, 177.533778)};
	\addlegendentry{Memo}
	%CYK
	\addplot [domain=0:3000,blue] {-1.787+0.00678267*x+0.0000610278*x^2+2.85858e-8*x^3};
	%Memo
	\addplot [domain=0:10000,blue,samples=100] {0.0530078 +0.00295173*x+2.00198e-6*x^2-5.47402e-11*x^3};
	\addplot [red,smooth] coordinates {
	  (0, 0.021556)
		(50, 0.415)
		(100, 1.645444)
		(150, 3.863778)
		(200, 23.074278)
		(250, 34.742611)
		(300, 54.224722)
		(350, 84.801)
		(400, 106.520556)
		(450, 167.9285)
		(500, 378.708389)
		(550, 1288.034833)};
\end{axis}
	\end{tikzpicture}
	\end{minipage}
	}
\caption{Recursive definition of Sierpinski graphs $T_n$ for $n>0$ (left) and running time (in ms) of different parsers analyzing $T_n$ for varying values of~$n$ (right).}
\label{f:timing-sierpinski}
\end{figure}

Running time of the three parsers has been measured for Sierpinski
graphs $T_n$ for different values of~$n$. Each $T_n$ consists of $2n+1$
triangles. $T_0$ is just a single triangle, and $T_n$ (for $n>0$) is
made of $T_k$, $T_m$, and $T_{n-k-m-1}$ as shown in
\figref{f:timing-sierpinski} where $k = \lfloor (n-1)/3 \rfloor$ and $m =
\lfloor (n-k-1)/2 \rfloor$, i.e., the $2n+1$ triangles of $T_n$ are as
equally distributed to $T_k$, $T_m$, and $T_{n-k-m-1}$ as possible.

\figref{f:timing-sierpinski} shows the running time of the different
parsers applied to $T_n$ with varying value~$n$,
measured on an iMac 2017, 4.2 GHz Intel Core i7, OpenJDK 12.0.1 with
standard configuration, and is shown in milliseconds on the $y$-axis
while $n$ is shown on the $x$-axis. Note the substantial speed-up when
using memoization (called ``Memo'' in the legend) compared to the plain
GPRS parser (called ``GPSR''). In fact, the GPSR parser
using memoization allows to parse Sierpinski graphs which cannot be
parsed in practice by the other two parsers. Moreover, maintaining the
memoization store is insignificant with respect to memory
consumption: The memoization store grows to $7n+2$ pairs when
parsing~$T_n$, i.e., memoization adds only linear space requirements.

Moreover, we reconsider the examples of series-parallel graphs and
structured flowcharts, which we have used in~\cite{hoffmann-minas:19}:

\def\elit{\lb e^{xy}}
\def\Glit{\lb G^{xy}}
\def\Gilit{\lb G^{xz}}
\def\Giilit{\lb G^{zy}}
  The following rules
  generate series-parallel graphs \cite[p.~99]{habel:92}:
  $$
    \startLit \too_0 \Glit \qquad
              \Glit \too_1  \elit \qquad
              \Glit \too_2 \Glit \, \Glit \qquad
              \Glit \too_3 \Gilit \, \Giilit 
  $$

\begin{figure}[bt]
	\hspace*{\fill}
	\includegraphics[scale=0.7]{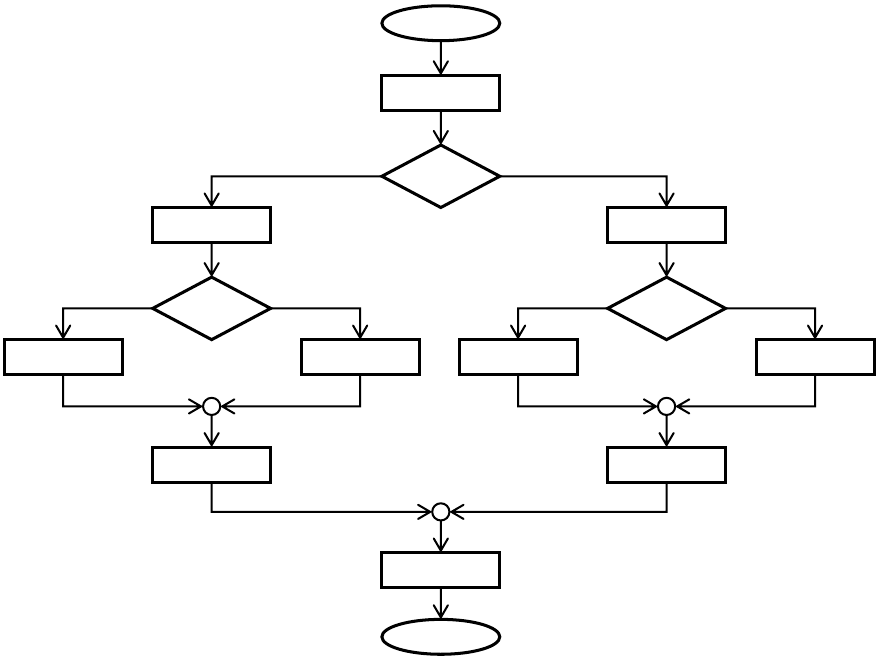}
	\hspace*{\fill}
	\includegraphics[scale=0.7]{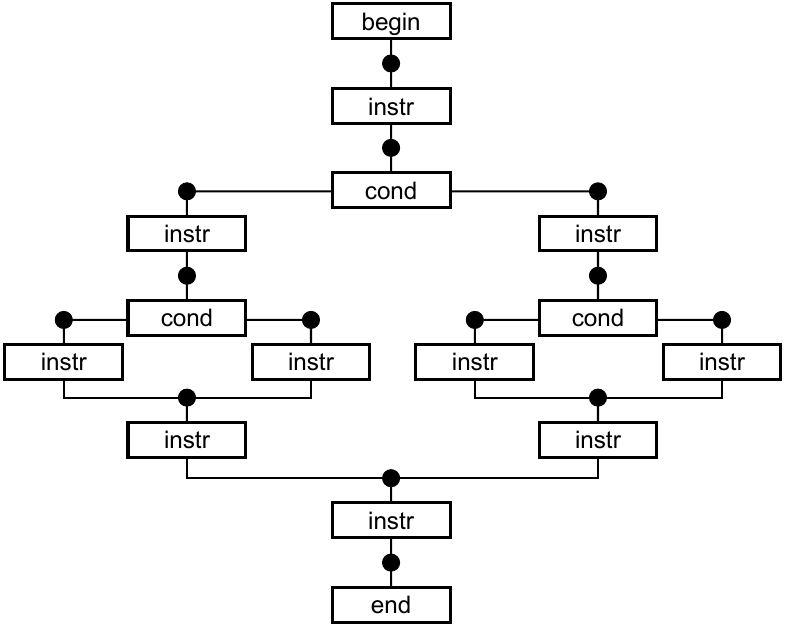}
	\hspace*{\fill}
	\caption{A structured flowchart 
	         and its graph representation.}
   	\label{f:sfc}
\end{figure}

Structured flowcharts are flowcharts that do not allow arbitrary jumps,
but represent structured programs with conditional statements and while
loops. They consist of rectangles containing instructions, diamonds
that indicate conditions, and ovals indicating begin and end of the
program. Arrows indicate control flow; see \figref{f:sfc} for an
example (text within the blocks has been omitted). Flowcharts are easily represented by graphs as also shown
in \figref{f:sfc}. The following rules generate all graphs representing
structured flowcharts:
\begin{eqnarray*}
	\startLit  &\to& \lb{begin}^x \, \lb P^{xy} \, \lb{end}^y \\
	\lb P^{xy} &\to& \lb S^{xy} \, \mid \, 
			 \lb P^{xz} \, \lb S^{zy} \\
	\lb S^{xy} &\to& \lb{instr}^{xy} \, \mid \,
 \lb{cond}^{xuv} \, \lb P^{uy} \, \lb P^{vy} \, \mid \, 
			 \lb{cond}^{xuy} \, \lb P^{ux} 
	\end{eqnarray*}%
%
%\end{example}
%
None of these grammars is PSR because their CFAs have conflicts. We
used these examples in~\cite{hoffmann-minas:19} to compare GPSR parsers
with CYK parsers. We extend these experiments here and additionally
compare these parsers with a GPSR parser using memoization.

As in~\cite{hoffmann-minas:19}, we employ GPSR parsers with two
different strategies for series-parallel graphs and for structured
flowcharts. GPSR~1 employs a breadth-first
strategy whereas GPSR~2 applies a more sophisticated strategy. It
requires grammar rules to be annotated with either first or second
priority. The GPSR~2 parser for series-parallel graphs gives rule~3
(series) precedence over rule~2 (parallel) whereas the GPSR~2 parser
for structured flowcharts gives sequences priority over conditional
statements.

\begin{figure}[bt]
	\begin{minipage}[b]{0.45\textwidth}
		\centering
		\begin{tikzpicture}[x=8mm,y=2.5mm,baseline=(0.base)]
			\node[anchor=east] (x) at (0,3) {$S_n = \mbox{}$};
			\node[o] (0) at (0,3) {};
			\node[o] (1a) at (1,0) {};
			\node[o] (1b) at (1,2) {};
			\node[o] (1c) at (1,4) {};
			\node[o] (1d) at (1,6) {};
			\node[o] (2a) at (2,0) {};
			\node[o] (2b) at (2,2) {};
			\node[o] (2c) at (2,4) {};
			\node[o] (2d) at (2,6) {};
			\node[o] (6a) at (3,0) {};
			\node[o] (6b) at (3,2) {};
			\node[o] (6c) at (3,4) {};
			\node[o] (6d) at (3,6) {};
			\node (3a) at (3.5,0) {$\cdots$};
			\node (3b) at (3.5,2) {$\cdots$};
			\node (3c) at (3.5,4) {$\cdots$};
			\node (3d) at (3.5,6) {$\cdots$};
			\node[o] (4a) at (4,0) {};
			\node[o] (4b) at (4,2) {};
			\node[o] (4c) at (4,4) {};
			\node[o] (4d) at (4,6) {};
			\node[o] (5) at (5,3) {};
			\node (c1) at (0.5,6) {\small$1$};
			\node (c2) at (1.5,7) {\small$2$};
			\node (c3) at (2.5,7) {\small$3$};
			\node (cn) at (4.5,6) {\small$n$};
				\path (0) edge[->] (1a) edge[->] (1b) edge[->] (1c) edge[->] (1d);
				\path (5) edge[<-] (4a) edge[<-] (4b) edge[<-] (4c) edge[<-] (4d);
				\path (1a) edge[->] (2a);
				\path (1b) edge[->] (2b);
				\path (1c) edge[->] (2c);
				\path (1d) edge[->] (2d);
				\path (2a) edge[->] (6a);
				\path (2b) edge[->] (6b);
				\path (2c) edge[->] (6c);
				\path (2d) edge[->] (6d);
		\end{tikzpicture}
					\caption{Definition of series-parallel graphs $S_n$.}
		\label{f:S-n}
	\end{minipage}
	\hfill
	\begin{minipage}[b]{0.5\textwidth}
		\centering
		\includegraphics[width=0.95\textwidth]{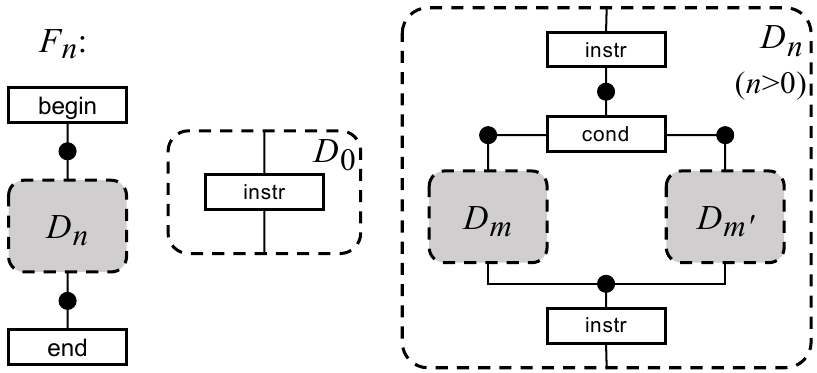}
		\caption{Definition of flowchart graphs $F_n$.}
		\label{f:sfc-rec}
  	\end{minipage}
\end{figure}

Running time of the parsers has been measured for series-parallel
graphs $S_n$ as shown in \figref{f:S-n} and for flowcharts $F_n$
defined in \figref{f:sfc-rec}. Each $F_n$ consists of $n$ conditions
and $3n+1$ instructions. The flowchart in \figref{f:sfc} is in fact
$F_3$. $F_n$ has a subgraph $D_n$, which, for $n>0$, contains subgraphs
$D_m$ and $D_{m'}$ with $n = m + m' + 1$. Note that the conditions in
$F_n$ form a binary tree with $n$ nodes when we ignore instructions. We
always choose $m$ and $m'$ such that it is a complete binary tree.
These shapes $S_n$ and $F_n$ turned out to be typical for
series-parallel graphs and flowcharts. Other shapes that linearly grow
with a parameter~$n$ show comparable results and could have been used
instead.

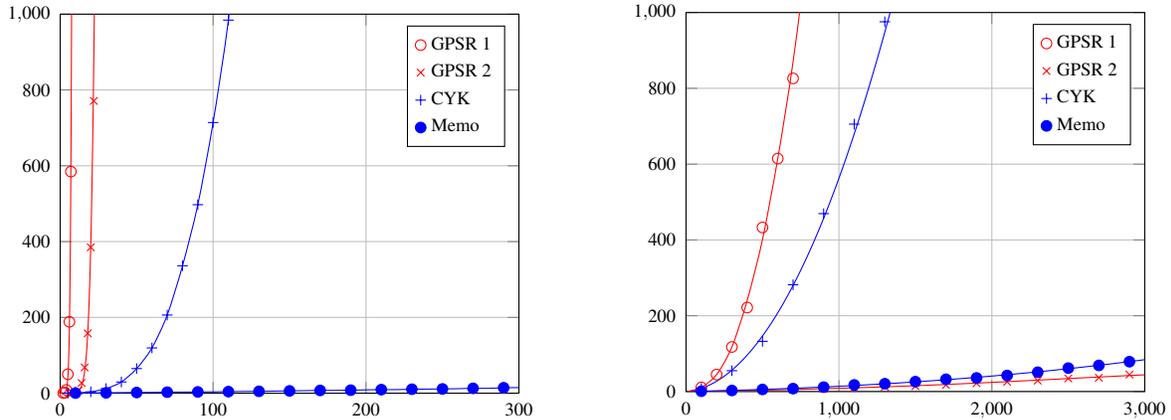
\begin{figure}[t]
	\begin{minipage}{0.48\textwidth}
		% !TEX root = main.tex
%
	\begin{tikzpicture}
	\begin{axis}[
	  width=\textwidth, 
	  xmin=0, xmax=300, 
	  ymin=0.1, ymax=1000, 
	  legend style={legend pos=north east}
	  ]
	% Cmd line: de.unibwm.inf2.grappa.examples.sp.SPTest -g sp2 -graph serial4 -stat -pt -steps -s 1 -w -r 12 -top 2 10
	\addplot [mark=o,red,only marks] coordinates {
		(1, 0.0615)
		(2, 0.2942)
		(3, 1.7925)
		(4, 8.6106)
		(5, 50.0811)
		(6, 188.4217)
		(7, 584.7055)
		(8, 1660.8867)
		(9, 4287.7453)
		(10, 10169.5684)};
	\addlegendentry{GPSR 1}
	%  Cmd line: de.unibwm.inf2.grappa.examples.sp.SPTest -g opt_sp2 -graph serial4 -stat -pt -steps -s 2 -w -r 12 -top 2 28
	\addplot [mark=x,red,only marks] coordinates {
		(2, 0.0533)
		(4, 0.0896)
		(6, 0.2419)
		(8, 0.6902)
		(10, 2.2027)
		(12, 7.0806)
		(14, 27.0448)
		(16, 68.0163)
		(18, 157.9009)
		(20, 384.8701)
		(22, 771.0757)
		(24, 1505.8055)
		(26, 2882.2717)
		(28, 5218.0045)};
	\addlegendentry{GPSR 2}
	% Cmd line: cyk.sp.SPTest -graph serial4 -stat -pt -steps -s 10 -w -r 15 -top 2 150
	\addplot [mark=+,blue,only marks] coordinates {
		% (1, 0.033807)
		% (2, 0.04492)
		% (3, 0.063145)
		% (4, 0.093966)
		(5, 0.135251)
		% (6, 0.183566)
		% (7, 0.249859)
		% (8, 0.338137)
		% (9, 0.454466)
		(10, 0.529231)
		(20, 3.582692)
		(30, 12.249231)
		(40, 29.658692)
		(50, 64.970154)
		(60, 119.416)
		(70, 206.261615)
		(80, 336.212154)
		(90, 497.303769)
		(100, 714.098154)
		(110, 984.298462)
		(120, 1351.446077)
		(130, 1750.311077)
		(140, 2270.030308)
		(150, 2906.737154)};
	\addlegendentry{CYK}
	% Cmd line: de.unibwm.inf2.grappa.examples.sp.SPTest -g sp2_memo -graph serial4 -stat -pt -steps -s 10 -w -r 50 -top 2 150
	\addplot [mark=*,blue,mark repeat=2,only marks] coordinates {
		(10, 0.33725)
		(20, 0.666)
		(30, 0.989)
		(40, 1.469)
		(50, 1.800604)
		(60, 2.079292)
		(70, 2.675042)
		(80, 2.886937)
		(90, 3.388312)
		(100, 3.565479)
		(110, 4.218979)
		(120, 4.668562)
		(130, 4.947042)
		(140, 5.451583)
		(150, 6.087062)
		(160, 6.698271)
		(170, 7.311708)
		(180, 7.604563)
		(190, 8.000375)
		(200, 8.862833)
		(210, 9.430042)
		(220, 9.461979)
		(230, 10.484583)
		(240, 10.814646)
		(250, 11.296854)
		(260, 11.901292)
		(270, 12.898646)
		(280, 12.899875)
		(290, 14.387875)
		(300, 14.752854)};
	\addlegendentry{Memo}
\addplot [domain=0:10,red,samples=100] {exp(-4.70481+1.84704*x-0.0163261*x^2-0.00297511*x^3)};
\addplot [domain=0:30,red,samples=100] {exp(-3.93141+0.360466*x+0.0164985*x^2-0.000487597*x^3)};
\addplot [blue,smooth] coordinates {
	(1, 0.033807)
	(2, 0.04492)
	(3, 0.063145)
	(4, 0.093966)
	%(5, 0.135251)
	%(6, 0.183566)
	%(7, 0.249859)
	%(8, 0.338137)
	%(9, 0.454466)
	(10, 0.529231)
	(20, 3.582692)
	(30, 12.249231)
	(40, 29.658692)
	(50, 64.970154)
	(60, 119.416)
	(70, 206.261615)
	(80, 336.212154)
	(90, 497.303769)
	(100, 714.098154)
	(110, 984.298462)
	(120, 1351.446077)
	(130, 1750.311077)
	(140, 2270.030308)
	(150, 2906.737154)};
	\addplot [blue,smooth] coordinates {
		(10, 0.33725)
		(20, 0.666)
		(30, 0.989)
		(40, 1.469)
		(50, 1.800604)
		(60, 2.079292)
		(70, 2.675042)
		(80, 2.886937)
		(90, 3.388312)
		(100, 3.565479)
		(110, 4.218979)
		(120, 4.668562)
		(130, 4.947042)
		(140, 5.451583)
		(150, 6.087062)
		(160, 6.698271)
		(170, 7.311708)
		(180, 7.604563)
		(190, 8.000375)
		(200, 8.862833)
		(210, 9.430042)
		(220, 9.461979)
		(230, 10.484583)
		(240, 10.814646)
		(250, 11.296854)
		(260, 11.901292)
		(270, 12.898646)
		(280, 12.899875)
		(290, 14.387875)
		(300, 14.752854)};
\end{axis}
	\end{tikzpicture}
	\end{minipage}
	\hfill
	\begin{minipage}{0.48\textwidth}
		% !TEX root = main.tex
%
	\begin{tikzpicture}
	\begin{axis}[
	  width=\textwidth, 
	  xmin=0, xmax=3000, 
	  ymin=0.1, ymax=1000, 
	  legend style={legend pos=north east}
	  ]
	% Cmd line: de.unibwm.inf2.grappa.examples.structured_fc.SFDTest -g lr -graph rec -pt -s 100 -r 12 -stat -pt -steps -w -top 2 1100
	\addplot [mark=o,red,only marks] coordinates {
	  (100, 10.8963)
		(200, 44.8455)
		(300, 117.6238)
		(400, 221.6062)
		(500, 432.8635)
		(600, 614.7939)
		(700, 826.2087)
		(800, 1160.1856)
		(900, 1636.7894)
		(1000, 2366.6428)
		(1100, 3088.6889)};
	\addlegendentry{GPSR 1}
	%  Cmd line: de.unibwm.inf2.grappa.examples.structured_fc.SFDTest -g opt_lr -graph rec -pt -s 100 -r 10 -stat -pt -steps -w 3000
	\addplot [mark=x,red,mark repeat=2,only marks] coordinates {
	  (100, 0.5458)
		(200, 1.1086)
		(300, 1.8428)
		(400, 2.6593)
		(500, 3.7888)
		(600, 4.392)
		(700, 5.3548)
		(800, 6.3847)
		(900, 7.3158)
		(1000, 9.5687)
		(1100, 10.3258)
		(1200, 11.379)
		(1300, 12.4042)
		(1400, 13.5828)
		(1500, 14.203)
		(1600, 19.546)
		(1700, 17.3154)
		(1800, 19.8606)
		(1900, 21.9091)
		(2000, 24.3713)
		(2100, 25.5129)
		(2200, 30.2986)
		(2300, 28.8561)
		(2400, 31.5084)
		(2500, 34.536)
		(2600, 37.8033)
		(2700, 36.4734)
		(2800, 40.6572)
		(2900, 44.7099)
		(3000, 42.3836)};
	\addlegendentry{GPSR 2}
	% Cmd line: cyk.sfd.SFDTest -graph rec -pt -s 100 -r 10 -stat -pt -top 1 -steps -w 2200
	\addplot [mark=+,blue,mark repeat=2,only marks] coordinates {
	  (100, 8.277778)
		(200, 23.956222)
		(300, 55.060556)
		(400, 85.734778)
		(500, 132.479667)
		(600, 204.381)
		(700, 281.740333)
		(800, 354.918889)
		(900, 469.435444)
		(1000, 546.562111)
		(1100, 705.489444)
		(1200, 814.057111)
		(1300, 975.559)
		(1400, 1105.173444)
		(1500, 1272.593333)
		(1600, 1391.145889)
		(1700, 1653.129333)
		(1800, 1822.314778)
		(1900, 2057.824444)
		(2000, 2163.462778)
		(2100, 2564.649667)
		(2200, 2763.215222)};
	\addlegendentry{CYK}
	% Cmd line: de.unibwm.inf2.grappa.examples.structured_fc.SFDTest -g memo_lr -graph rec -pt -s 100 -r 10 -stat -pt -steps -w 3000
	\addplot [mark=*,blue,mark repeat=2,only marks] coordinates {
	  (100, 0.8619)
		(200, 1.7952)
		(300, 2.8507)
		(400, 4.2259)
		(500, 5.5821)
		(600, 6.2612)
		(700, 7.7652)
		(800, 10.7439)
		(900, 11.6983)
		(1000, 12.6181)
		(1100, 17.4841)
		(1200, 19.5021)
		(1300, 20.6382)
		(1400, 25.4238)
		(1500, 26.0919)
		(1600, 28.0125)
		(1700, 32.4026)
		(1800, 30.9297)
		(1900, 35.8769)
		(2000, 41.3391)
		(2100, 42.4399)
		(2200, 46.9807)
		(2300, 51.0111)
		(2400, 58.6557)
		(2500, 61.8721)
		(2600, 66.4378)
		(2700, 68.9241)
		(2800, 73.9047)
		(2900, 78.8268)
		(3000, 83.3371)};
	\addlegendentry{Memo}
\addplot [domain=10:3000,red,samples=100] {0.00712929*x - 5.3542e-7*x^2 + 2.49603e-9*x^3 - 
	4.91845e-13*x^4};
	\addplot [domain=10:3000,blue,samples=100] {0.00907212*x + 3.8549e-6*x^2 + 1.09746e-9*x^3 - 9.18945e-14*x^4};
	\addplot [domain=10:2500,blue,samples=100]  {2.03194e-8*x^3+0.00050897*x^2+0.0335308*x};
	\addplot [red,smooth] coordinates {
		(10, 0.1982)
		(100, 10.8963)
		(200, 44.8455)
		(300, 117.6238)
		(400, 240.6062)
		(500, 400.8635)
		(600, 614.7939)
		(700, 870.2087)
		(800, 1200.1856)
		(900, 1700.7894)
		(1000, 2266.6428)
		(1100, 3000.6889)};
	\end{axis}
	\end{tikzpicture}
	\end{minipage}
	\caption{Running time (in ms) of different parsers analyzing series-parallel graphs $S_n$ (left) and structured flowcharts $F_n$ (right) with varying value $n$.}
	\label{f:timing-sp-sfd}
\end{figure}

\figref{f:timing-sp-sfd} shows the running time of the different
parsers applied to $S_n$ and $F_n$ with varying value~$n$ on the same
platform as for Sierpinski graphs. The experiments again show that the
GPSR parser with memoization is substantially faster than the CYK
faster and even more faster than most of the GPSR parsers. Only GPSR~2
for structured flowcharts is a bit faster than the memoization parser
because it need not maintain the memo store. But note that realizing
the hand-tailored strategy for the GPSR~2 parser required additional
programming work, whereas the memoization parser has been generated by
the \textsc{Grappa} distiller without any further manual work.

Maintaining the memoization store when parsing series-parallel graphs
and structured flowcharts is insignificant with respect to memory
consumption, just as for Sierpinski graphs: The memoization store grows
to $38n-39$ (for $n > 2$) and $18n +4$ pairs when parsing~$S_n$ and
$F_n$, respectively, i.e., memoization adds only linear space
requirements.
% !TEX root = main.tex

\section{Conclusions}\label{s:concl}

We have proposed to use memoization to make GPSR parsing faster by
memorizing nonterminal edges that have been created in 
the search process and that are discarded by plain GPSR parsing
although this information could be reused later. Our experiments with
three example languages (Sierpinski graphs, series-parallel graphs, and
structured flowcharts) have shown that GPSR parsing with memoization is
in fact substantially faster and does not increase memory 
consumption significantly for theses examples. However, memoization
is not a silver bullet. It cannot speed up GPSR parsing when analyzing
invalid input graphs. In these cases, they must completely traverse the
entire search space, essentially falling back to plain GPSR parsing.
The same applies if one is not only interested in one successful parse,
but in all parses if the input graph is ambiguous.

Memoization techniques have also been used to speed up GLR parsers for
strings; J.R.~Kipps improved the original GLR algorithm from
$O(n^{k+1})$ where $k$ is the length of the longest rule to $O(n^3)$ 
using memoization~\cite{Kipps:1991}. And this
speed-up is independent of the input string being valid or
invalid.\footnote{As a matter of fact, in his thesis,
G.R.~Economopoulos~\cite[p.~184]{Economopoulos:2006} questions the
correctness of this claim because Kipps's parsing algorithm does not
terminate on grammars with hidden-left recursion, similar to Tomita's
original GLR parsing algorithm~\cite{Tomita:85}.} But memoization for
GLR parsing differs entirely from memoization for GPSR parsers proposed
here: A GLR parser searches for all parses of the input graph in
parallel, and all these ``parsing processes'' are synchronized by
reading one input string token after the other. Memoization helps to
speed up reduce steps in the graph-structured stack. A GPSR parser,
instead, must try different ``reading sequences'' of the input graph,
and memoization helps to reuse information that has been found earlier
in a different reading sequence. Kipps's memoization approach in fact
resulted in a parsing algorithm with the same running time complexity
as the newer BRNGLR parsers~\cite{Scott:07}, which do not need
memoization at all. But parsing for HR grammars is in general NP-hard.
So there cannot be a general fast parsing algorithm for HR grammars, and one must
depend on techniques like memoization, as suggested in this paper, to
obtain efficient parsers.

In future work, we will apply GPSR parsing with memoization to examples
from natural language processing, in particular for parsing
\emph{Abstract Meaning Representations} (AMR)~\cite{Banarescu.etAl:13}.
PSR parsing cannot be applied there because almost all grammars are
ambiguous in this field. In particular, we would like to compare our
parser with the state of the art in this field, i.e., the Bolinas
parser~\cite{chiang-et-al:2013} by D.~Chiang, K.~Knight \emph{et al.}\
that implements the polynomial algorithm for HR grammars devised
in~\cite{lautemann:90} and the s-graph parser
\cite{Groschwitz-Koller-Teichmann:15} by A.~Koller \textsl{et al.}

\bibliographystyle{eptcs}
\bibliography{refs}
\end{document}